\documentclass[11pt]{article}

\usepackage{cite}
\usepackage{latexsym}
\usepackage{amsmath,amsfonts,amssymb}
\usepackage{accents}

\newcommand{\be}{\begin{equation}}
\newcommand{\ba}{\begin{eqnarray}}
\newcommand{\ee}{\end{equation}}
\newcommand{\ea}{\end{eqnarray}}

\newcommand{\barr}{\begin{array}}
\newcommand{\ear}{\end{array}}
\newcommand{\ns}{\normalsize}

\newcommand{\ax}{\alpha}
\newcommand{\bx}{\beta}
\newcommand{\cx}{\gamma}
\newcommand{\dx}{\delta}

\newcommand{\ox}{\omega}

\newcommand{\Ox}{\Omega}

\newcommand{\ap}{\alpha'}

\newcommand{\um}{\undertilde m}

\newcommand{\ubx}{\undertilde \bx}
\newcommand{\ucx}{\undertilde \cx}
\newcommand{\udx}{\undertilde \dx}

\newcommand{\wg}{\wedge}
\newcommand{\del}{\partial}

\newcommand{\rep}[1]{\mathbf{#1}}

\newcommand{\nn}{\nonumber}

\newcommand{\tox}{\tilde\omega}

\newcommand{\CT}{\kappa} 
\newcommand{\K}{\Upsilon} 
\newcommand{\conn}{w} 

\newcommand{\cK}{\mathcal{K}}

\newcommand{\cG}{\mathcal{G}}

\newcommand{\cZ}{\mathcal{Z}}

\numberwithin{equation}{section}

\begin{document}

\begin{titlepage}

\title{
   \hfill{\normalsize arXiv:0709.1932}
    \vskip 2cm
   {\Large\bf Heterotic String Compactifications on Half-flat Manifolds
   II}\\[0.5cm]} 
   \setcounter{footnote}{0}
\author{
  {\ns\large
    Sebastien Gurrieri$^1$\footnote{email: sebgur@gmail.com}}~,
  {\ns\large Andr\'e Lukas$^2$\footnote{email: lukas@physics.ox.ac.uk}}~
  {\ns and Andrei Micu$^3$\footnote{email: amicu@th.physik.uni-bonn.de}
    $^{,}$\footnote{On leave from IFIN-HH Bucharest.}}
  \\[0.5cm]
{\it\ns $^1$ Kansai Furansu Gakuin}\\
  {\ns 536-1 Waraya-cho Marutamachi dori Kuromon Higashi iru,}\\
  {\ns  Kamigyo-ku, 602-8144 Kyoto, Japan} \\[0.5cm]
  {\it\ns $^2$Rudolf Peierls Centre for Theoretical Physics, University of
   Oxford}\\
 {\ns 1 Keble Road, Oxford OX1 3NP, UK} \\[.5cm]
 {\it\ns $^3$Physikalisches Institut der Universit\"at Bonn}\\
 {\ns Nussallee 15, 53115, Bonn, Germany}}

\date{}

\maketitle

\begin{abstract}
  In this paper, we continue the analysis of heterotic string
  compactifications on half-flat mirror manifolds by including the
  10-dimensional gauge fields.  It is argued, that the heterotic Bianchi
  identity is solved by a variant of the standard embedding. Then, the
  resulting gauge group in four dimensions is still $E_6$ despite the fact
  that the Levi-Civita connection has ${\rm SO}(6)$ holonomy.  We derive the
  associated four-dimensional effective theories including matter field terms
  for such compactifications. The results are also extended to more general
  manifolds with ${\rm SU}(3)$ structure.
\end{abstract}

\thispagestyle{empty}

\end{titlepage}


\section{Introduction}

In recent attempts to stabilise moduli in string compactifications fluxes have
played a crucial role. It has also been realised that the notion of flux can
be generalised to include geometric fluxes which can be described in terms of
manifolds with restricted structure group. In this paper we will concentrate
on six-dimensional manifolds with ${\rm SU}(3)$ structure which are the
nearest cousins of Calabi--Yau manifolds. There exists a further
generalisation to non-geometric fluxes which are related to backgrounds with
${\rm SU}(3) \times {\rm SU}(3)$ structure, but in this paper we will stay in
the realm of geometric compactifications.

In the context of heterotic string compactifications, manifolds with ${\rm
  SU}(3)$ structure play an important role. Soon after Calabi--Yau
compactifications were proposed in Ref.~\cite{CHSW} it was realised that in the
presence of $H$-flux, the supersymmetric ground state of the heterotic string
corresponds to an internal manifold which is complex, but non-K\"ahler
\cite{AS,CH}. More recently, in
Refs.~\cite{BD1}-\cite{GPR} such compactifications were classified in terms of
  $\mathrm{SU}(3)$ structures 
which is the natural way to approach this problem. It was argued in
Refs.~\cite{BBDP,CCDL,BD2} that in this way a  
superpotential is generated and some of the Calabi--Yau moduli get
fixed.

This mechanism for moduli stabilisation is one of the most attractive phenomenological
features of ${\rm SU}(3)$ structure compactifications, particularly in the context of
the heterotic string, where only one type of conventional flux, NS-NS flux, is available.
For Calabi-Yau compactifications, this stabilises the complex structure moduli
\cite{RW} but not the K\"ahler moduli (and the dilaton) which remain flat
directions. The only know way to generate a perturbative superpotential for
the K\"ahler moduli is indeed to use manifolds with ${\rm SU}(3)$ structure. In
Ref.~\cite{GLM} the superpotential for a particular class of such ${\rm
  SU}(3)$ structure manifolds, so-called half-flat mirror manifolds, was
derived and subsequently analysed in Ref.~\cite{dCGLM}. However, the analysis
was restricted to the gravitational sector (zeroth order in $\alpha'$) and the
gauge/matter sector was not addressed in detail. A detailed analysis turned
out to be quite involved due to the large number of terms which appear in the
reduction of a 10-dimensional gauge theory to four dimensions.

In this paper, we will show how to overcome these difficulties, using the
heterotic Gukov formula for the superpotential (for the original version of the
formula in the context of type IIA and M theory see
Refs.~\cite{GVW,SG}). Based on this approach we will 
explicitly calculate the four-dimensional effective theory including the gauge
field sector for heterotic compactifications on half-flat mirror manifolds and
their generalisations.  This result will allow us to address a number of
questions which have been the main motivation behind this work. What is the
four-dimensional low-energy gauge group for such non Calabi-Yau
compactifications? What is the four-dimensional (gauge matter) particle
spectrum? Do any of these four-dimensional gauge matter fields pick up masses
from the (geometrical) fluxes? Is the low-energy gauge group spontaneously
broken?

Let us explain in more detail how we proceed in deriving the four-dimensional
effective theory. One of the main obstacles to overcome in heterotic string
compactifications is to find a solution to the Bianchi identity for the NS-NS
field strength $H$. Since our knowledge about manifolds with ${\rm SU}(3)$
structure is fairly limited explicitely constructing bundles over such spaces
is an ambitious task (for recent developments see
Refs.~\cite{CL,FY}). Nevertheless we always have the standard embedding at 
our disposal where the background gauge field is set equal to the spin
connection. It represents the simplest solution to the Bianchi identity and
will be adopted in this paper. To determine the expansion of the
10-dimensional fields into low-energy modes we will be guided by an "adiabatic
principle" which has been proposed and successfully applied to type II string
theory in Ref.~\cite{GLMW,GM} and has been shown in Ref.~\cite{GLM} to lead to
a consistent description of the gravitational sector in heterotic theories. The
basic assumption underlying this principle is that half-flat mirror manifolds
(and their generalisations) can be considered as "perturbations" of Calabi-Yau
manifolds and, hence, that their low-energy spectrum is identical to the one
of the associated Calabi-Yau manifolds. We will show that this principle leads
to a consistent description of the heterotic gauge field sector. In
particular, we will show that the low-energy gauge group remains $E_6$ in
agreement with the adiabatic principle.  This conclusion may be surprising
since the Levi-Civita connection for manifolds with ${\rm SU}(3)$ structure
has, in general, a holonomy group $SO(6)\simeq SU(4)$ which suggests the
low-energy gauge group $SO(10)$. This is, in fact, what has been proposed in
Ref.~\cite{GLM}. Here we will show that $E_6$ is the correct low-energy gauge
group.

Having decided upon the expansion of 10-dimensional fields into low-energy
modes we will use the heterotic Gukov formula for the superpotential
(gravitino mass)
\begin{equation}
  \label{W}
  e^{K/2} W = \frac{e^{\phi}}{\sqrt 2 \cK ||\Ox||} \int \Omega \wedge (H + i
  dJ ) \; ,
\end{equation}
which was derived from first principles in Ref.~\cite{GLM}. It provides a way
of computing the K\"ahler potential $K$ and superpotential $W$ of the low-energy
theory in terms of the NS-NS flux $H$ and the forms $\Omega$ and $J$ which
characterise the ${\rm SU}(3)$ structure. For half-flat mirror manifolds, both
$H$ and $dJ$ are non-zero at zeroth order in $\ap$ and this leads to a
superpotential which is linear in the K\"ahler moduli \cite{GLM}.  At first
order in $\ap$ the above formula receives a contribution from the Chern-Simons
correction to $H$. Given the expansion of the gauge fields we can explicitely
compute the Chern-Simons term and integrate to obtain the superpotential
including matter field terms.

It has been known since the early work in Ref.~\cite{witten} that the
definition of the K\"ahler moduli in terms of the 10-dimensional fields is
modified at first order in $\ap$ by a certain combination of the matter
fields. This aspect of compactifications with matter fields often
significantly complicates the task of finding the correct definitions of the
low-energy fields. In our context this problem can be quite elegantly dealt
with.  Due to the existence of a K\"ahler-moduli dependent superpotential at
zeroth order in $\ap$ one expects the matter field corrections to appear at
order $\ap$ in Eq.~\eqref{W}.  This is indeed what we will find.  The correct
definition of the moduli can then be read off from these additional terms.
Based on these ideas we will carry out the full reduction for half-flat mirror
manifolds~\cite{GLMW,GLM} and then extend the results to the
generalised half-flat manifolds introduced in Refs.~\cite{ferrara,GLW}.

The paper is organised as follows. We start with a review of heterotic
Calabi--Yau compactifications based on the standard embedding.  This is mostly
to fix conventions and as a reference point for the more involved calculation
in the half-flat case. Then, in Section \ref{sec:hf}, we perform the analogous
analysis for half-flat mirror manifolds.  In Section \ref{sec:hfbi}, we first
discuss the Bianchi identity, in Section \ref{sec:hfW} we compute the
gravitino mass and, finally, in Section \ref{sec:KW}, we find the associated
four-dimensional effective theory. For the correct interpretation of the
result it will prove useful to include $H$-flux which we will do in
Section~\ref{sec:Hflux}. Then, in Section \ref{sec:genhf}, we extend the
results of Section \ref{sec:hf} to generalised half-flat manifolds. Finally,
in Section \ref{sec:con}, we conclude and present future directions for
research. Formulae and conventions for Calabi-Yau manifolds and the group
$E_8$ have been collected in the Appendix.

\section{Warm up: Calabi--Yau compactifications of the heterotic string}
\label{sec:CY}

In this section we perform the compactification of the heterotic string on
Calabi--Yau manifolds using the standard embedding. Since this material is
well-known we will keep the discussion brief and we refer the reader to the
standard textbooks, for example \cite{GSW}, for a more detailed treatment.
However, in view of our earlier discussion, we will perform the computation in
the gauge matter sector in an unusual way, using the Gukov formula \eqref{W}.

\subsection{The spectrum in 10 and 4 dimensions}

We start with the low energy action of the heterotic string in 10 dimensions
which is given by supergravity coupled to a super-Yang-Mills theory with gauge
group $E_8 \times E_8$.\footnote{In this paper we focus on $E_8 \times E_8$
  and do not discuss the $SO(32)$ case.} The bosonic spectrum -- which is what
we are mostly interested in -- is given by the graviton $G_{MN}$, the
antisymmetric tensor field $B_{MN}$ and the dilaton $\phi$ in the gravity
sector and by the gauge fields $A_M^I$, where $I$ is an adjoint index of $E_8
\times E_8$ and runs from $1, \ldots, 496$. The action for these fields is
given by 
\begin{equation}
  \label{s10}
  S =- \frac12 \int e^{-2 \phi} \left[ R * \mathbf{1} - 4 d \phi \wedge * d
  \phi + \frac12 H \wedge * H + \ap \left({\rm Tr} F\wedge *F - {\rm tr}
  \tilde R^2 \right) \right] \; ,
\end{equation}
where ${\rm tr} \tilde R^2$ stands for the Gauss-Bonet combination. Here we
have kept the dependence on $\ap$ which is going to be a useful expansion
parameter. We will neglect any terms at order $\ap^2$ or higher. The field
strengths $F$ and $H$ are defined by
\begin{equation}
  \label{Fc}
  F = d A + A \wedge A \; ,
\end{equation}
and
\begin{equation}
  \label{Hc}
  H = dB + \ap (\omega_L - \omega_{\mathrm{YM}}) \; .
\end{equation}
Here $\omega_L$ and $\omega_{\mathrm{YM}}$ are the Chern-Simons three-forms
\begin{eqnarray}
  \label{CSGr}
  \omega_L & = & {\rm tr} \big(\tilde R \wedge \tilde \conn - \frac13 \tilde
  \conn \wedge \tilde \conn \wedge \tilde \conn \big) \; ,\\
  \omega_{\mathrm{YM}} & = & {\rm Tr} \big( F \wedge A - \frac13 A \wedge A
  \wedge A \big) \; .\label{CSYM}
\end{eqnarray}
The modified spin connection one-form, $\tilde \conn$, is given in 
terms of the Levi-Civita connection, $\conn$, by \cite{BdR}
\begin{equation}
  \label{modsc}
  \tilde \conn_{MN}{}^P = \conn_{MN}{}^P - \frac12 H_{MN}{}^P \; ,
\end{equation}
and all curvature tensors denoted by $\tilde R$ are computed in terms of this
modified connection. Finally, the symbol ${\rm Tr}$ above denotes $1/30$ of
the trace in the adjoint of $E_8 \times E_8$. 

Taking the exterior derivative of Eq.~\eqref{Hc} one obtains the well-known
Bianchi identity
\begin{equation}
  \label{BI}
  dH = \ap \big( {\rm tr} \tilde R \wedge \tilde R - {\rm Tr} F \wedge F
  \big)\; . 
\end{equation}
The simplest solution to this equation -- which we will also adopt in this
paper -- is the so called standard embedding where the background gauge field
(for one of the $E_8$ group factors) is set equal to the spin connection
$\tilde \conn$. In the absence of $H$-fluxes the connection has ${\rm SU}(3)$
holonomy and this choice of background breaks the gauge group to $E_6 \times
E_8$. Here and in the rest of the paper we shall always have a "hidden sector"
$E_8$ gauge factor which we will often omit from the discussion. The gauge
matter fields which reside in four-dimensional chiral multiplets arise from
the internal components of the gauge fields and consist of $h^{2,1}$
$\rep{27}$-plets and $h^{1,1}$ $\rep{\overline{27}}$-plets where $h^{1,1}$ and
$h^{2,1}$ denote the Hodge numbers of the Calabi--Yau manifold. Therefore the
net number of families is given by $|h^{1,1} - h^{2,1}| = |\chi|/2$.

In addition to the fields discussed above there are a number of gravitational
fields. Apart from the four-dimensional metric tensor in the gravity multiplet
we have $h^{1,1}$ (complexified) K\"ahler moduli, $t^i$, and $h^{2,1}$ complex
structure moduli, $z^a$, as well as the axio-dilaton $s$, all of them in
four-dimensional chiral multiplets.  Finally we have the so called "bundle
moduli" which parameterise deformations of the gauge bundle. In this paper we
will not be concerned with this last class of fields and we will ignore them.

\subsection{Four-dimensional effective action}

Let us move on to the four-dimensional action for the fields described above.
It will be useful to organise the discussion according to the order in $\ap$
at which different terms appear. Let us start with order zero. At this order,
only the compactification of ten dimensional supergravity contributes and
leads to the four-dimensional supergravity sector coupled to the moduli
fields. The action is given by \footnote{For our conventions, see Appendix
  \ref{conv}.}
\begin{equation}
  \label{modkt}
  S_{0,{\rm  kinetic}} =- \int\left[ \frac12 R *\! {\bf 1} +d \phi\wg *d\phi
  +\frac14 e^{4\phi} da \wg *da + g_{ij}d t^i\wg *d \bar t^j\nn + g_{a \bar b}
  dz^a \wg *d \bar z^b \right]
\end{equation}
where $t^i$ denote the complexified K\"ahler moduli. They are obtained by
expanding the complexified K\"ahler form $B + i J$ into two-forms $\omega_i$,
\begin{equation}
  \label{tdef}
  B + i J = (b^i + i v^i) \ox_i \equiv t^i \ox_i \; ,
\end{equation}
with $i,j,\ldots =1,\dots ,h^{1,1}$.  Further, $z^a$ denote the complex
structure moduli introduced in Eq.~\eqref{csmoduli}, $\phi$ is the
four-dimensional dilaton and its partner $a$, the universal axion, is the
dual of the four-dimensional tensor field $B_{\mu \nu}$. The kinetic terms in
\eqref{modkt} can be obtained from the usual zeroth order K\"ahler potential,
$K_0$, given by
\begin{equation}
  K_0 = K_{K} + K_{cs} + K_{s} \, ,
  \label{K0}
\end{equation}
with
\begin{eqnarray}
  K_{\rm cs} & = & -\ln\left( \cK ||\Omega||^2\right) \; ,\\[2mm]
  K_K & = & -\ln \cK \; ,\\[2mm]
  K_s & = & - \log (i(\bar s - s)) \label{Ks}\; ,
\end{eqnarray}
and
\begin{equation}
  \label{s}
  s = a +ie^{-2\phi} \; ,
\end{equation}
Here, $\cK$ denotes the Calabi--Yau volume \eqref{vol} and $\Omega$ is the
holomorphic $(3,0)$ form. Explicit expressions for $K_K$ and $K_{\rm cs}$ are
given in Appendix \ref{CYconv}.  At zeroth order in $\ap$, the superpotential
$W_0$ vanishes and, hence, the above K\"ahler potential completely specifies the
theory at this order.

Let us now discuss the ten-dimensional gauge fields and their descendants in
four dimensions which appear at first order in $\ap$.  For the standard
embedding case the massless matter fields can be obtained from expanding the
internal gauge-fields in $(0,1)$ harmonic forms with values in the holomorphic
and anti-holomorphic tangent bundle.  On a Calabi-Yau manifold these spaces
are known to be isomorphic to the cohomology groups 
$H^{2,1}(X)$ and $H^{1,1}(X)$ and we can, therefore, write
\begin{eqnarray}
  \label{chfdef}
  A_{\bar \ax} &=& A^{(0)}_{\bar \ax} +A^{(1)}_{\bar \ax}\\
  A^{(1)}_{\bar \ax}&=&||\Omega||^{-1/3} (\omega_i)_{\bar
    \ax}{}^{\bar \bx} e_{\bar \bx}^{\bar \ubx} {\bar T}_{\bar \ubx \bar P}
    C^{i \bar P} + ||\Omega||^{1/3} (\eta_a)_{\bar \ax}{}^{\bx}
    e_{\bx}^{\ubx} T_{\ubx P} D^{a P} \; . \label{chfdef1}
\end{eqnarray}
Here $C^{i \bar P}$ and $D^{aP}$ denote the matter fields in the
$\rep{\overline{27}}$ and $\rep{27}$ respectively, $\omega_i$ are the harmonic
$(1,1)$ forms introduced earlier and the rank two symmetric tensors $\eta_a$
are defined in 
terms of the $(2,1)$ forms, see Eq.~\eqref{eta}. By $A^{(0)}$ we have denoted
the background gauge field and $A^{(1)}$ contains the matter field
fluctuations around this background. Having chosen the standard embedding, the
background is set equal to the Calabi--Yau spin-connection, that is
\begin{equation}
  \label{bgA}
  A^{(0)}_{\bar \ax} = \conn_{\bar \ax \bx \bar \cx} S^{\bx \bar \cx} \; ,
\end{equation}
where $S^{\bx \bar \cx}$ are the generators of ${\rm SU}(3)\subset E_8$,
defined in Appendix~\ref{E8}.  Note that the indices $\bx$ and $\bar \cx$
should in principle be tangent space indices. We have glossed over this
subtlety in the above formula in order not to overload the notation. In
\eqref{chfdef} however, the nature of indices is important, especially when
taking derivatives, and therefore we have explicitely included the vielbeins
to convert between curved and tangent space indices. Note the factors
$||\Omega||^{\pm 1/3}$ in the expansion \eqref{chfdef1} which correspond to a
particular definition of the gauge matter fields $C^{i \bar P}$ and $D^{a P}$.
As we shall see below these factors are required in order to make the
superpotential holomorphic.  Note that $||\Omega||$ does not depend on the
Calabi-Yau coordinates and, therefore, these additional factors do not
complicate the calculation.

The kinetic terms for the gauge matter fields can be easily obtained by
calculating the field strength (specifically the components with one internal
and one external leg) associated to Eq.~\eqref{chfdef1} and inserting the
result into the Yang-Mills part of the 10-dimensional action~\eqref{s10}. One
finds 
\begin{equation}
  \label{kinCD}
  S_{1,{\rm  kinetic}} = -\ap \int \left[ 4 g_{ij} ||\Omega||^{-2/3} dC^i_P
  \wedge *d \bar C^{j P} + g_{a \bar b} ||\Omega||^{2/3} d \bar D^{aP} \wedge
  * dD^{\bar b}_P \right] \; .
\end{equation}
These kinetic terms can be obtained from the matter field K\"ahler potential
\begin{equation}
  \label{KCD0}
  K_1 = 4 \ap g_{ij} ||\Omega||^{-2/3}  C^i_P \bar C^{jP} + \ap
  g_{a \bar b} ||\Omega||^{2/3} D^{aP} \bar D^{\bar b}_P \; ,
\end{equation}
which should be added to the zeroth order result~\eqref{K0}. Up to constant
normalisations this is precisely the K\"ahler potential computed in
Ref.~\cite{DKL}, which confirms the expansion \eqref{chfdef1}.

The scalar potential for the gauge matter fields can be calculated by
computing the purely internal components of the field strength associated to
Eqs.~\eqref{chfdef}, \eqref{chfdef1} and inserting the result into the
Yang-Mills part of the 10-dimensional action~\eqref{s10}.  One may then
attempt to read off the superpotential and the $D$-terms from the result.
However, as discussed earlier, there is a simpler and cleaner way of deriving
the superpotential based on the heterotic Gukov formula \eqref{W}. Let us
follow this route and see how the known result for the matter field
superpotential can be reproduced.

In the case presently under discussion, that is Calabi-Yau compactifications
in the absence of $H$-flux, no superpotential arises at zeroth order in $\ap$
as is clear from the Gukov formula~\eqref{W}.  However, at first order in
$\ap$, $H$ can still pick up a purely internal component due to the
Chern-Simons term in Eq.~\eqref{BI}. Therefore, using Eqs.~\eqref{Hc} ,
\eqref{CSGr} and \eqref{CSYM} with the gauge field Ansatz~\eqref{chfdef}
inserted, the Gukov formula~\eqref{W} can be written as
\begin{equation}
  \label{W1}
  W =  \ap \int \Omega \wedge (\omega_L-\omega_{YM}) = - \ap \int \Omega
  \wedge {\rm Tr} \left( F \wedge A - \frac13 A \wedge A \wedge A \right)
  ^{(1)}\; . 
\end{equation}
In this equation, the pure Yang-Mills background contribution due to $A^{(0)}$
and its field strength
\begin{equation}
  F^{(0)} = d A^{(0)} + A^{(0)} \wedge A^{(0)} \;.
\end{equation}
is canceled by the Lorentz Chern-Simons term by virtue of the standard
embedding.  It should, therefore, be omitted from the expression on the RHS of
Eq.~\eqref{W1} which is indicated by the superscript $(1)$. Hence, only terms
at least linear in $A^{(1)}$ or its field strength contribute. In addition,
due to the presence of $\Omega$ in Eq.~\eqref{W1} only the $(0,3)$ piece of
the Chern-Simons term is relevant. Therefore only the $(0,2)$ part of $F$,
$F_{(0,2)}$, enters the calculation. Since we have expanded the gauge fields
in $(0,1)$ harmonic forms with values in the (anti)holomorphic tangent bundle
the part of $F_{(0,2)}$ linear in $A^{(1)}$ vanishes. Let us see more
explicitly how this works. The terms which are linear in the matter fields can
schematically be written as
\begin{equation}
  F^{(1)} = d A^{(1)} + 2\big[ A^{(0)},A^{(1)} \big] \; .
\end{equation}
In the first term the derivative can act on the forms $\ox_i$ and $\eta_a$ or
on the vielbeins.  When the derivatives act on the forms then, due to
antisymmetrisation, we end up with exterior derivatives, which vanish in the
Calabi--Yau case.  The remainder takes the form
\begin{eqnarray}
  (F^{(1)})_{\bar \ax \bar \bx} & = & ||\Omega||^{1/3} (\eta_a)_{\bar
    \ax}{}^{\cx} 
    \left[ \nabla_{\bar \bx} e_{\cx}{}^{\ucx} T_{\ucx P} + \conn_{\bar \bx
    \ubx \bar \ucx} [T_{\udx P} , S^{\ubx \bar \ucx}]  e_\cx{}^{\udx} \right]
    - (\bar \ax \leftrightarrow \bar \bx ) + ... \nn \\
    & = & ||\Omega||^{1/3} (\eta_a)_{\bar \ax}{}^{\cx} \left[\nabla_{\bar \bx} 
    e_{\cx}{}^{\ucx} - \conn_{\bar \bx \udx}{}^{\ucx} e_\cx{}^{\udx} \right]
    T_{\ucx P}- (\bar \ax \leftrightarrow \bar \bx ) + ... \; ,
\end{eqnarray}
where we have focused on terms containing $\eta_a$ and the dots stand for
analogous terms containing $\ox_i$ . In the last equality we have used the
commutator \eqref{cr8327}. Note that the second term in this commutation
relation does not contribute as the spin connection is a ${\rm SU}(3)$ Lie
algebra valued one-form and therefore its contraction with the metric
vanishes. In the bracket we recognise the defining equation for the spin
connection in terms of the vielbein and, hence, the above terms vanish.  The
same conclusion holds for the terms containing $\ox_i$ and, hence, $F^{(1)}$ is
zero. Thus we have indeed shown that $(0,2)$ part of the field strength
originates from the commutator term, that is
\begin{equation}
  \label{F02CY}
  F_{(0,2)} = A^{(1)} \wedge A^{(1)} \; ,
\end{equation}
or in components
\begin{equation}
  \label{F02CYc}
  F_{\bar \ax \bar \bx} = \left[||\Omega||^{-1/3} (\ox_i)_{\bar \ax}{}^{\bar
  \cx} \bar T_{\bar \cx \bar P} C^{i \bar P} + ||\Omega||^{1/3} (\eta_a)_{\bar
  \ax}^{\cx} T_{\cx P} D^{a P} , ||\Omega||^{-1/3} (\ox_j)_{\bar \bx}{}^{\bar
  \dx} \bar T_{\bar \dx \bar R} C^{j \bar R} + ||\Omega||^{1/3} (\eta_b)_{\bar
  \bx}^{\dx} T_{\dx R} D^{b R} \right] \; .
\end{equation}
Based on this result, let us now perform a similar analysis for the full
combination of Chern-Simons terms in \eqref{W1}. We have already mentioned
that the pure background part in this combination cancels between the gravity
and gauge field terms due to the standard embedding. Linear terms in $A^{(1)}$
cannot be present as they would lead to gauge non-invariant terms in the
superpotential. Hence, we are left with quadratic and cubic terms in
$A^{(1)}$. However, using \eqref{F02CY}, it is easy to see that the terms
quadratic in $A^{(1)}$ cancel in Eq.~\eqref{W1}. Therefore we can write the
superpotential for the charged fields as
\begin{equation}
  \label{W1fin}
  W = - \ap \frac23 \int \Omega \wedge {\rm Tr}\left( A^{(1)} \wedge A^{(1)}
  \wedge A^{(1)} \right) \; . 
\end{equation}
Substituting the expression \eqref{chfdef1} for $A^{(1)}$ we obtain the
final result for the order $\ap$ superpotential, $W_1$, which
reads\footnote{In the context of heterotic M-theory a similar method for
  deriving this superpotential was recently used in \cite{Schmidt}.}
\begin{equation}
  \label{WCD}
  W_1 = -\frac13 \ap \left[\bar j_{\bar P \bar R \bar S} \cK_{ijk} C^{i
  \bar P} C^{j \bar R} C^{k \bar S} + j_{PRS} \tilde \cK_{abc} D^{a P} D^{bR}
  D^{cS} \right] \; .
\end{equation}
Here we have used the trace relation \eqref{jdef} and $\cK_{ijk}$ and
$\tilde\cK_{abc}$ are the triple intersection numbers defined in Eqs.
\eqref{tin} and \eqref{mtin}.  This is the well-known cubic superpotential
which we have derived from the heterotic Gukov formula \eqref{W}.

In order to compute the $N=1$ supergravity potential we also need the
$D$-terms. Given that we know the matter field content and the K\"ahler
potential, these can, of course, be calculated purely from $N=1$ supergravity
\cite{nillesrev} 
\begin{equation}
  \label{Dsugra}
  D^x = G_I (T^x)^I{}_J \xi^J \; .
\end{equation}
Here, $G_I$ are the derivatives of the supergravity $G$-function, $G= K +
\mathrm{ln} |W|^2$, with respect
to the matter fields $\phi^I$ and $T^x$ denote the gauge group generators in
the representation in which $\xi_I$ transforms.
However, as a useful consistency check, the $D$-terms can also be directly
obtained from 10 dimensions using the formula~\cite{Iman}
\begin{equation}
  \label{Dterm}
  D^x = \int F^x \wg * J \; ,
\end{equation}
which is similar to the Gukov formula~\eqref{W} for the superpotential.  Here
$x$ is an $E_6$ adjoint index. Since $J$ is a $(1,1)$-form we need the $(1,1)$
components of $F$ to evaluate the expression on the RHS.  They can be
calculated similar to the $(0,2)$ component of $F$ above, the main difference
being that antisymmetrisation of the indices does not lead to exterior
derivatives so terms with derivatives acting on forms no longer vanish.
Instead, we have
\begin{eqnarray}
  \label{F11}
  F_{\ax \bar \bx}  & = & F^{(0)}_{\ax \bar \bx} + ||\Omega||^{-1/3}
  \nabla_{\ax}  
  (\omega_i)_{\bar \bx}{}^{\bar \cx} \bar T_{\bar \cx \bar P} C^{i \bar P} +
  ||\Omega||^{1/3} \nabla_{\ax} (\eta_a)_{\bar \bx}{}^\cx T_{\cx P} D^{aP} \nn
  \\ 
  & & -||\Omega||^{-1/3} \nabla_{\bar \bx} 
  (\omega_i)_{\ax}{}^{\cx} T_{\cx P} \bar C^{i P} - ||\Omega||^{1/3}
  \nabla_{\bar \bx} (\bar \eta_{\bar a})_{\ax}{}^{\bar \cx} \bar T_{\bar \cx
    \bar P} \bar D^{\bar a \bar P}+ \big [A^{(1)} ,A^{(1)} \big] \; .
\end{eqnarray}
Contributions to the $D$-terms can only arise from terms which involve $E_6$
generators and, hence, recalling the commutation relations \eqref{cr327b},
only from the last term in the above expression.  Performing the integral in
\eqref{Dterm} we obtain
\begin{equation}
  \label{DCD}
  D^x = 4 ||\Omega||^{-2/3} g_{ij} C^{i \bar P} \bar C^{j R} k^x_{\bar P R} 
  - ||\Omega||^{2/3} g_{a \bar b} D^{a P} \bar D^{\bar b \bar R} k^x_{P \bar
  R} \; ,
\end{equation}
which is precisely what one obtains from the supergravity formula
\eqref{Dsugra} applied to \eqref{KCD0}.
This ends our review of heterotic Calabi--Yau compactifications with standard
embedding. 

\section{Compactification on half-flat mirror manifolds}
\label{sec:hf}

In this section, we will compactify the heterotic string, including its gauge
field sector, on half-flat mirror manifolds~\cite{GLMW}, following the
strategy outlined in the previous section. Due to the lack of explicit
constructions for these manifolds, a rigorous derivation is not really
possible. Instead we will adopt the "adiabatic approach" of
Ref.~\cite{GLMW,GM}  
which is based on the assumption that these new manifolds can be seen as
"small" variations of the underlying Calabi--Yau manifold. Many of the
standard Calabi--Yau methods can then be transferred to half-flat mirror
manifolds.  This approach has been successfully applied to type II mirror
symmetry~\cite{GLMW} and to the gravitation sector of the heterotic
string~\cite{GLM}.

Let us be more explicit about this approximation and point out some of its
consequences. As it will become clear in the next section we will introduce
parameters $e_i$ to quantify the departure form ordinary Calabi--Yau
manifolds. Therefore one of the main assumption is that these parameters are
small and terms which contain more than two powers of $e_i$ will be neglected.
By the adiabatic approach we will still consider a ten-dimensional background
metric which is a product of a four-dimensional Minkowski space and the metric
on the internal manifold. For this to be a solution of the ten-dimensional
supersymmetry equations, the manifold with $\mathrm{SU}(3)$ structure has to
obey certain conditions \cite{CCDLMZ,BBDG,FL,MPZ}. We do not impose these
conditions on the compactification manifolds in the first place as our purpose
here is only to obtain a low energy effective action. The condition we should
however require for a consistent result, is that the (supersymmetric) minima
of the (super)potential we find here should be at points in the moduli space
where the ten-dimensional equations are indeed satisfied. We will deal with
these conditions elsewhere \cite{dCLM}.

After a brief review of manifolds with ${\rm SU}(3)$ structure and half-flat
mirror manifolds in the next sub-section, we will start by finding a solution
to the Bianchi identity~\eqref{BI} for half-flat mirror manifolds which is
analogous to the standard embedding. Then, we will evaluate the heterotic
Gukov formula~\eqref{W} for half-flat mirror manifolds. In the pure Calabi-Yau
case, evaluating this formula has given us information merely about the matter
field superpotential but not the K\"ahler potential, despite the explicit K\"ahler
potential dependence in Eq.~\eqref{W}. The reason is that the superpotential
in the standard Calabi-Yau case is of order $\ap$ and, hence, calculating the
matter field K\"ahler potential (which is also of order $\ap$) requires
evaluating Eq.~\eqref{W} at order $\ap^2$. Terms at this order are beyond the
scope of our calculation and, therefore, we resorted to a standard reduction
of the Yang-Mills action to determine the K\"ahler potential from the matter
field kinetic terms.  For half-flat mirror manifolds, on the other hand, the
torsion flux generates a superpotential at zeroth order in $\ap$. As a result,
the Gukov formula at order $\ap$ will provide us with information about both
the superpotential and the matter field K\"ahler potential.  In addition, we
will be able to infer another crucial piece of information, namely the correct
definition of the K\"ahler moduli fields, which receive order $\ap$ matter field
corrections, as is well-known for the Calabi-Yau case~\cite{witten}.
Generalising our set-up further by adding $H$-flux then allows us to find an
analogous correction to the definition of the complex structure moduli. As we
will see, this provides us with sufficient information to completely determine
the four-dimensional gauge matter field action at order $\ap$.

\subsection{Half-flat SU(3) structure manifolds}
\label{sec:revhf}

Before we proceed with the computation, we review some of the required
properties of manifolds with ${\rm SU}(3)$ structure and the specific
sub-class of half-flat (mirror) manifolds (for a more formal description of
manifolds with $\mathrm{SU}(3)$ structure see for example \cite{CS}). Manifolds
with ${\rm SU}(3)$ 
structure are almost complex manifolds for which the structure group of the
frame bundle reduces to ${\rm SU}(3)$. They can be described in terms of an
invariant two-form $J$ (the fundamental form) and an invariant three-form
$\Omega$ which is of type $(3,0)$ with respect to the almost complex
structure. Manifolds with ${\rm SU}(3)$ structure can be classified according
to their intrinsic torsion $\tau$, which is associated to the connection which
preserves the structure (that is, which annihilates the forms $J$ and
$\Omega$).  The intrinsic torsion is a one-form taking values in ${\rm
  su}(3)^\perp$ where
\begin{equation}
  \label{so6dec}
  {\rm so}(6) = {\rm su}(3) \oplus {\rm su}(3)^\perp \; .
\end{equation}
Here ${\rm so}(6)\sim{\bf 15}$ and ${\rm su}(3)\sim{\bf 8}$ denote the Lie
algebras of ${\rm SO}(6)$ and ${\rm SU}(3)$, respectively, and ${\rm
  su}(3)^\perp\sim{\bf 3}\oplus{\bf\bar{3}}\oplus{\bf 1}$ is the part
perpendicular to ${\rm su}(3)$.  Unlike for Calabi-Yau manifolds, the forms
$J$ and $\Omega$ are no longer closed and the expressions for $dJ$ and
$d\Omega$ can be used to read off the intrinsic torsion $\tau$ and, hence, to
characterise the manifold. Half-flat manifolds are formally defined by
imposing additional restrictions on $dJ$ and $d\Omega$, namely
\begin{equation}
  \label{hfdef}
  d(J \wedge J) = d\, \mathrm{Im} \Omega = 0 \; .
\end{equation}
These remove half of the torsion components which is why these manifolds are
sometimes also called half-integrable.

For practical purposes it is more useful to define a class of half-flat
manifolds starting from underlying Calabi--Yau manifolds \cite{GLMW}. As
mentioned earlier, this "adiabatic" approach has the advantage of providing a
fairly explicitly framework for calculations with many of the standard
Calabi-Yau techniques applicable.  One starts by postulating the existence of
a set of two-forms $\ox_i$ and a symplectic set of three-forms
$(\ax_A,\bx^B)$, where $(\ax_A)=(\ax_0,\ax_a)$ and $(\ax^B)=(\ax^0,\ax^b)$.
They are of course the analogue of the harmonic two and three forms on a
Calabi-Yau manifold and still satisfy the standard normalisation relations
\eqref{norm2} and \eqref{norm3}. However, they are no longer closed but
instead satisfy the identities
\begin{eqnarray}
  \label{hfalg}
  d \ox_i & = & e_i \beta^0 \; , \qquad d \tox^i =0 \; , \\
  d \ax_0 & = & e_i \tox^i \; , \qquad d \ax_a = d \bx^A = 0 \; , \nn
\end{eqnarray}
where $e_i$ are torsion flux parameters.  Apart from this modification, the
properties of Calabi-Yau manifolds listed in Appendix~\ref{CYconv} are assumed
to remain valid. In particular, the moduli are defined by expanding the ${\rm
  SU}(3)$ invariant forms $J$ and $\Omega$ into the above sets of forms, as in
Eqs.~\eqref{Kms} and \eqref{csms}. Then, it is easy to see that the first of
the half-flat conditions \eqref{hfdef} is implied by the primitivity of the
three forms $(\ax_A,\bx^A)$ on a Calabi--Yau manifold.  The second condition
is satisfied because the standard choice ${\cal Z}^0=1$ implies that ${\rm
  Im}\Omega$ does not contain $\ax_0$ which is the only non-closed three-form.
We will refer to half-flat manifolds with the above set of forms and
properties as half-flat mirror manifolds, due to their appearance in the
context of type II mirror symmetry with NS-NS flux~\cite{GLMW}.

The heterotic string on such half-flat mirror manifolds at zeroth order in
$\ap$ has been discussed in Ref.~\cite{GLM} and the results can be easily
summarised.  Since most of the standard Calabi-Yau relations still hold it
follows that the moduli K\"ahler potential~\eqref{K0} remains unchanged
from the Calabi-Yau case.  The only modification at zeroth order in $\ap$ is
the appearance of a superpotential~\cite{GLM}
\begin{equation}
  \label{Whf}
  W_0 = e_i t^i \; ,
\end{equation}
for the K\"ahler moduli $t^i$, as can be easily seen from Eq.~\eqref{W}.
For later purposes it will be useful to explicitely derive the components
of the intrinsic torsion from the relations \eqref{hfalg}. It is not hard to
show \cite{GLMW} that
\begin{equation}
  \label{tau}
  \begin{aligned}
    \tau_{\ax \bx}{}^{\bar \cx}  & = \frac{1}{4 ||\Omega||^2} (e_i
    \tox^i)_{\ax \bx \bar \ax \bar \bx} \Omega^{\bar \ax \bar \bx \bar \cx}
    \;, \\[1mm]
    \tau_{\ax \bx}{}^\cx & = - \frac{i}{2} (e_i v^i) (\beta^0)_{\ax
    \bx}{}^\cx \; ,
  \end{aligned}
\end{equation}
with components other than the complex conjugate of the above being zero.  It
is important to note that the first two indices of the torsion are of the same
complex type. Further, primitivity of $\beta^0$ implies the contraction of the
torsion tensor with $J$ vanishes, that is
\begin{equation}
  \label{tJ0}
  \tau_{mnp} J^{np} = 0 \; .
\end{equation}
Hence, the torsion tensor has no components in the singlet part of ${\rm
  su}(3)^\perp\sim{\bf 3}\oplus{\bf\bar{3}}\oplus{\bf 1}$.  The contorsion
tensor, $\CT$, which we shall also need later on, can be written in terms of
the torsion tensor as
\begin{equation}
  \label{TCT}
  \CT_{mnp} = \tau_{mnp} + \tau_{pmn} + \tau_{pnm} \; .
\end{equation}
Eqs.~\eqref{tau} and \eqref{tJ0} imply that the singlet part of $\CT$ also
vanishes.  It is also worth pointing out that the internal components of the
field strength $H$ are non-zero for half-flat mirror manifolds, even in the
absence of genuine NS-NS flux.  This happens because the forms $\omega_i$ are
no longer closed and, hence, taking the exterior derivative of the $B$ field
Ansatz~\eqref{tdef} together with Eq.~\eqref{hfalg} one finds, apart from the
usual terms involving four-dimensional derivatives, that
\begin{equation}
  \label{Hhf}
  H = e_i b^i \beta^0 \; .
\end{equation}

\subsection{Solving the Bianchi identity}
\label{sec:hfbi}

In order to apply the Gukov formula~\eqref{W} we need to compute the field
strength $H$ from its definition~\eqref{Hc}. This, in turn, requires finding a
background gauge field configuration which satisfies the Bianchi identity
\eqref{BI}.  Here, we would like to discuss the simplest possibility, that is
a gauge field background obtained from a standard embedding. However, things
are not quite so straightforward as we have various connections available to
set the gauge field equal to. The most immediate choice seems to be to set the
gauge field equal to the Levi-Civita connection, $\conn$, of the half-flat
mirror manifold.  There are two obvious problems with such a choice. Firstly,
it is the modified connection $\tilde\conn = \conn-H/2$ which enters the
curvature term in the Bianchi identity. While this made no difference in the
Calabi-Yau case, it does here since, as we have seen in Eq.~\eqref{Hhf}, the
internal part of $H$ is non-vanishing. Therefore, setting the gauge field
equal to $\conn$ means the Bianchi identity is not strictly satisfied. 
Secondly, and perhaps more importantly,
$\conn$ (and presumably $\tilde\conn$ as well) has holonomy ${\rm SO}(6)$,
leading to a gauge symmetry breaking to ${\rm SO}(10)$ rather than $E_6$. In
fact, such a breaking to ${\rm SO}(10)$ was predicted in Ref.~\cite{GLM}.
However, the adiabatic approach dictates that low-energy modes should be the
same as for the Calabi-Yau case and that, consequently, the gauge group should
be $E_6$. Such a breaking can be realised with the torsion connection
$\conn^{(T)}$ which has ${\rm SU}(3)$ holonomy. Schematically, it is related
to $\tilde\conn$ by
\begin{equation}
  \label{modsc1}
  \tilde \conn_m = \conn_m^{(T)} + \K_m^\parallel + \K_m^{\perp} \; ,
\end{equation}
where $\K_m^\parallel\in{\rm so}(3)$ and $\K_m^{\perp} \in{\rm so}(3)^\perp$,
and the tensor $\K$ will be explicitly determined below. This way of splitting up the
connection $\tilde\conn$ is in line with the possible steps for gauge symmetry breaking.
Specifically, the first two terms preserve an $E_6$
gauge group, while the further breaking to $\mathrm{SO}(10)$ is only due to
$\K_m^\perp$. We will show that this part can be absorbed into a
re-definition of the matter gauge fields $C$ and $D$.  The difference between
choosing $\tilde\conn$ and $\conn_m^{(T)} + \K_m^\parallel= \tilde \conn - \K_m^\perp$ therefore
amounts to a shift of the four-dimensional fields. Generally, one would
expect that a sensible choice of connection leads to a low-energy
supersymmetric vacuum with vanishing matter fields, $C=D=0$, and unbroken
gauge symmetry, while less suitable choices for the background may lead to
non-vanishing VEVs for $C$ and $D$ and symmetry breaking (or restoration).  In
keeping with the adiabatic approach we will set the gauge field equal to
$\conn_m^{(T)} + \K_m^\parallel$, so that the low-energy gauge group at this
stage is $E_6$.  As we will show, for this choice of background there is
indeed always a vacuum with $C=D=0$ and $E_6$ unbroken.
 
Let us try to make the above discussion more precise. We start by splitting
Lie-algebra valued forms into ${\rm su}(3)$ and ${\rm su}(3)^\perp$ parts
which we denote by superscripts $\parallel$ and $\perp$, respectively. The
torsion connection $\conn^{(T)}$ takes of course values in ${\rm su}(3)$
while, for the Levi-Civita connection, we can write~\footnote{Note that, 
  although the torsion $\tau$ is supposed to be an element of ${\rm
    su}(3)^\perp$ and, hence, $\tau^\parallel=0$, in general the parallel
  component of the con-torsion, $\CT^\parallel$ is non-zero. Indeed, from
  Eq.~\eqref{TCT}, we find $\CT_{\bar \ax \bx \bar \cx} = - \tau_{\bar \ax
    \bar \cx \bx} \ne 0$. Moreover, since the singlet part of $\CT^\perp$
  vanishes, this component of $\CT$ must be part of $\CT^\parallel$.}
\begin{equation}
  \label{su3con}
  \conn_m = \conn_m^{(T)} - \CT_m^{\parallel} - \CT_m^{\perp} \; .
\end{equation}
Similarly, we can also think of $H$ as an ${\rm so}(6)$ valued one-form which
can be decomposed as 
\begin{equation}
  \label{Hsplit}
  H_m = H_m^{\parallel} + H_m^{\perp} \; .
\end{equation}
As the three forms $\beta^A$ are primitive, the explicit form~\eqref{Hhf}
of $H$ implies that the singlet part in $H^\perp$ is also zero.

Since $H$ and $\CT$ appear on the same footing in the modified connection
$\tilde\conn$ in Eq.~\eqref{modsc} it is useful to introduce the notation
\begin{equation}
  \label{Kdef}
  \K_{mnp} = -\left(\CT_{mnp} + \frac12 H_{mnp} \right) \; ,
\end{equation}
which leads to Eq.~\eqref{modsc1}. Given the expressions~\eqref{Hhf} for $H$
and for the (con)torsion tensor for a half-flat mirror manifold,
Eqs.~\eqref{tau} and \eqref{TCT}, the tensor $\K^\parallel$ takes the form
\begin{equation}
  \label{K||}
  (\K_{\bar \ax}^\parallel)_{\ax \bar \bx} = - \frac12 (e_i t^i)
  (\beta^0)_{\bar \ax \ax \bar \bx} \; ,
\end{equation}
while, using Eq.~\eqref{bchi}, the orthogonal component can be written as
\begin{eqnarray}
  \label{K|-}
  (\K_{\bar \ax}^\perp)_{\ax \bx} \bar \Omega^{\ax \bx \cx} & = & -i \frac{e_i
    t^i}{\cK} g^{a \bar b} K_{\bar b} (\eta_a)_{\bar \ax}{}^\cx \; , \\
  (\K_{\bar \ax}^\perp)_{\bar \bx \bar \cx} \Omega^{\bar \bx \bar \cx \bar \dx}
    & = & \frac{e_i t^i}{\cK} v^i (\omega_i)_{\bar \ax}{}^{\bar \dx} \; . \nn
\end{eqnarray}
Earlier, in Eqs.~\eqref{chfdef} and \eqref{chfdef1}, we have split the
internal gauge field $A_m$ as $A_m=A_m^{(0)}+A_m^{(1)}$ into a background term
$A^{(0)}$ and fluctuation term $A^{(1)}$, which is linear in the gauge matter
fields $C$ and $D$.  Comparison of \eqref{K|-} with the Ansatz for $A^{(1)}$
shows that $\K^\perp$ can indeed be absorbed into a re-definition of the
matter fields $C$ and $D$, as claimed earlier.  This means, we can set the
background gauge field to $\conn_m^{(T)} + \K_m^\parallel$ instead of
$\tilde\conn$ and write for the gauge field Ansatz
\begin{equation}
 A_m=A_m^{(0)}+A_m^{(1)}\; , \label{A}
 \end{equation}
 with
\begin{equation}
  A^{(0)}_m = \left( \conn^{(T)}_{m \ax \bar \bx} + \K^\parallel_{m \ax \bar \bx}
  \right) S^{\ax \bar \bx} \; , \label{A0}
\end{equation}
and $A^{(1)}$ as in Eq.~\eqref{chfdef1}, but with $C$ and $D$ re-interpreted.
Here $S^{\ax\bar\bx}$ are the generators of ${\rm SU}(3)$ in the branching
$E_8\rightarrow {\rm SU}(3)\otimes E_6$ (see Appendix~\ref{E8}).  The
background $A^{(0)}_m$ now takes values in ${\rm su}(3)$ and, hence, the
low-energy gauge group is $E_6$, in line with the adiabatic approach.  Since
we are simply shifting $\K^\perp$ between background and fluctuations without
changing the total gauge field, there should be no problem with this
procedure. However, there is one practical difficulty. Given the
choice~\eqref{A0} for $A^{(0)}$, the background part of the Chern-Simons terms
in the definition of $H$, Eq.~\eqref{Hc}, does not cancel by itself.  Let us
look at the order of the remainder. The perpendicular part $\CT^\perp$ of the
torsion is linear in the torsion parameters $e_i$ and its contribution to the
Chern-Simons term is of ${\cal O}(e_i^2)$. The RHS of Eq.~\eqref{Hc} is
suppressed by $\ap$, so the resulting contribution to $H$ is of ${\cal O}(\ap
e_i^2)$. Inserting this contribution to $H$ into the Gukov formula~\eqref{W} a
non-vanishing contribution of ${\cal O}(\ap e_i^3)$ can arise from
multiplication with $dJ$ (which by itself is of order $e_i$).  This is two
powers higher in flux than the terms we keep in our calculation and will,
hence, be discarded.

To summarise this section, we have chosen the background gauge fields to be the
$\mathrm{su}(3)$-valued connection \eqref{A0} so that the resulting
four-dimensional gauge group is $E_6$. This can be viewed as a
standard embedding of $\tilde \conn$ into the gauge field but with the $E_6$ breaking part,
$\K_m^\perp$, of the connection being absorbed into a redefinition of the charged fields $C$ and
$D$. This also represents a solution to the Bianchi identity \eqref{BI} at
order $\mathcal{O}(\ap e_i)$ and therefore it constitutes a consistent
background at this order.

\subsection{Gravitino mass term at order $\ap$}
\label{sec:hfW}

Having fixed the gauge field Ansatz, we can now proceed and evaluate formula
\eqref{W} for half-flat manifolds at order $\ap$. However, we have to keep in
mind that our manifolds are no longer complex but almost complex only. This
means that the complex coordinate types of the field strength $F$ cannot be
simply obtained by taking holomorphic or anti-holomorphic derivatives of the
gauge field in complex coordinates. Rather, we should first re-write the gauge
field Ansatz in real coordinates, then differentiate to compute the field
strength in real coordinates and only afterwards project to complex coordinate
types.  In real coordinates, the gauge field Ansatz~\eqref{A}, \eqref{A0},
\eqref{chfdef1} reads
\begin{equation}
 A_m=A_m^{(0)}+A_m^{(1)} \; ,\label{Adec}
\end{equation}
with
\begin{eqnarray} 
 A^{(0)}_m &=& \left( \conn^{(T)}_{m \ax \bar \bx} + \K^\parallel_{m \ax \bar \bx}
  \right) S^{\ax \bar \bx} \label{Areal0}\\ 
 A^{(1)}&=&||\Omega||^{-1/3} (\omega_i)_m{}^n \left({\bar T}_{n \bar P}
  C^{i \bar P} + T_{n P} \bar C^{i P} \right)\\
  && + ||\Omega||^{1/3} \left[(\eta_a)_m{}^n T_{n P} D^{a P} + (\bar \eta_{\bar a})_m{}^n \bar T_{n \bar
  P} \bar D^{\bar a \bar P} \right]\; .\label{Areal1}
\end{eqnarray}
Here, we have adopted the convention that the antiholomorphic pieces of the
generators $T_{n P}$, corresponding to the ${\rm SU}(3)\otimes E_6$
representations $({\bf\bar{ 3}},{\bf 27})$ and $({\bf 3},{\bf \bar{27}})$,
vanish, that is
\begin{equation}
  \label{Tconv}
  T_{\bar \ax P} = \bar T_{\ax \bar P} = 0 \; .
\end{equation}
More generally, having to work in real coordinates means that we have to be
careful when comparing to the Calabi-Yau formulae in the previous section and
convert them into complex coordinates first.

Our first task is to calculate the internal part of the gauge field strength
$F$.  To focus our discussion, let us for a moment assume that the background
gauge field $A^{(0)}$ equals the ${\rm SU}(3)$ connection $\conn^{(T)}$, that
is, let us discard the $\K^\parallel$ piece in Eq.~\eqref{Areal0} for now.
Then, the computation of the field strength is very similar to the computation
we have already described for Calabi--Yau manifolds. In particular, we can use
the covariant derivative with torsion, $\nabla^{(T)}$, associated to
$\conn^{(T)}$, to re-write the exterior derivative as
\begin{equation}
  \label{dt}
  (dA)_{mn} = \nabla^{(T)}_m A_n - \nabla^{(T)}_n A_m + 2 \tau_{mn}{}^p A_p\; .
\end{equation}
This means, apart from the torsion term on the right hand side which we have
to subtract, the formula for the field strength should be the same as in the
Calabi--Yau case but with the ordinary covariant derivatives replaced by
torsion covariant derivatives. Then, finally, we also have to take into account
the effect of a non-vanishing $\K^\parallel$ in \eqref{Areal0}. As in the
Calabi-Yau case, it will be useful to organise terms according to their power
in the matter fields $C$ and $D$. Terms in $F$ related to $\K^\parallel$ will
be either pure background terms or linear in the matter fields. The pure
background terms are not particularly interesting for us. In the Gukov
formula~\eqref{W} they lead to background terms which cancel up to higher
order terms and to terms linear in the matter fields which should be zero as a
consequence of gauge invariance.  The linear, $\K^\parallel$ related terms in
$F$, on the other hand, only result from the commutator term in the definition
of the field strength and, hence, do not involve derivatives. These terms, as
in fact the whole commutator in the expression for the field strength, can be
easily computed without the detour to real indices. Hence, for now, we will
only write the general expressions for these commutator terms in order not to
overload the equations.  This understood, we find for the (internal) field
strength in real indices
\begin{eqnarray}
  \label{Freal}
  F_{mn} & = & F^{(0)}_{mn} + ||\Omega||^{-1/3} \nabla^{(T)}_m (\ox_i)_n{}^q T_{qP}
  \bar C^{i P} + ||\Omega||^{-1/3} \nabla^{(T)}_m (\ox_i)_n{}^q \bar T_{q \bar
    P} C^{i \bar P} - (m \leftrightarrow n) \nn \\
  & & + 2 ||\Omega||^{-1/3} \tau_{mn}{}^r (\ox_i)_r{}^q T_{qP} \bar C^{i P}
  + 2 ||\Omega||^{-1/3} \tau_{mn}{}^r (\ox_i)_r{}^q \bar T_{q \bar P} C^{i
    \bar P} \nn \\
  & & + ||\Omega||^{1/3} \nabla^{(T)}_m (\eta_a)_n{}^q T_{qP}
  D^{a P} + ||\Omega||^{1/3} \nabla^{(T)}_m (\bar \eta_{\bar a})_n{}^q \bar
  T_{q \bar P} \bar D^{\bar a \bar P} - (m \leftrightarrow n) \\
  & & + 2 ||\Omega||^{1/3} \tau_{mn}{}^r (\eta_a)_r{}^q T_{qP} D^{a P}
  + 2 ||\Omega||^{1/3} \tau_{mn}{}^r (\bar \eta_{\bar a})_r{}^q \bar T_{q \bar
    P} \bar D^{\bar a \bar P} \nn \\
  & & + \big[ A^{(1)}_m , A^{(1)}_n \big] + 2 \big[ \K^\parallel_m, A^{(1)}_n \big ] \nn
  \; , 
\end{eqnarray}
where $F^{(0)}$ denotes the background field strength computed from $A^{(0)}$
in Eq.~\eqref{Areal0}, and $A^{(1)}$ and $\K^\parallel$ are explicitly given
in Eqs.~\eqref{Areal1} and \eqref{K||}.

Having derived this result for the field strength one could follow the
"traditional" route and derive the four-dimensional scalar potential by
computing ${\rm tr} F^2$ (as well as $H^2$) and integrate over the internal
space. We will indeed compute a few selected terms in the scalar potential in
this way later, in order to check our results. However, given the complexity
of Eq.~\eqref{Freal} there is no doubt that the full calculation is rather
tedious and that reading off the correct definitions of superfields and the
superpotential from the result is likely to be difficult. For example,
integrating over the internal space in the presence of an arbitrary number of
$(2,1)$ and $(1,1)$ forms will lead to integrals which are non-standard even
in the Calabi-Yau case. Further, the background curvature $F^{(0)}_{mn}$
enters the calculation explicitly. Although, the Ricci tensors for manifolds
with ${\rm SU}(3)$ structure in general and half-flat manifolds in particular
have been computed in Refs.~\cite{math,tibra}, the results are fairly
complicated. At any rate, we would need those results for the somewhat unusual
connection $\conn^{(T)}+\K^\parallel$ which are not readily available.  To
circumvent these obstacles we would like to base our calculation on the
Gukov-formula~\eqref{W}, which provides direct information about the gravitino
mass $m_{3/2}=e^{K/2} W$. As we will see, with a bit more work, the
so-obtained result for $m_{3/2}$ can be disentangled and provides information
about the K\"ahler potential and superpotential.

As in the Calabi-Yau case, the superpotential at order $\ap$ arises entirely
from the $(0,3)$ part of the Chern-Simons combination in \eqref{Hc}.
Therefore, we only need to know the $(0,2)$ component, $F_{(0,2)}$, of the field
strength which can be derived by projecting the result \eqref{Freal} onto the
$(0,2)$ subspace. Note that all derivatives in Eq.~\eqref{Freal} are torsion
covariant derivatives which commute with the almost complex structure $J$.
Therefore, converting to complex indices is as straightforward as for normal
complex manifolds. The second observation is that due to Eq.~\eqref{Tconv},
many of the terms in \eqref{Freal} vanish, when written in complex
indices. With these facts in mind we find
\begin{eqnarray}
  \label{F02hf}
  F_{\bar \ax \bar \bx} & = & 2 || \Omega||^{-1/3} \nabla^{(T)}_{[\bar \ax}
  (\omega_i)_{\bar \bx]}{}^{\bar \cx} \bar T_{\bar \cx \bar P} C^{i \bar P} +
  2 ||\Omega||^{1/3} \nabla^{(T)}_{[\bar \ax} (\eta_a)_{\bar \bx]}{}^{\cx}
  T_{\cx P} D^{a P} \nn \\
  & & + 2 ||\Omega||^{-1/3} \tau_{\bar \ax \bar \bx}{}^{\cx}
  (\omega_i)_\cx{}^\dx T_{\dx P} \bar C^{i P} + 2 ||\Omega||^{1/3} \tau_{\bar
    \ax \bar \bx}{}^{\cx} (\bar \eta_{\bar a})_\cx^{\bar \dx} \bar T_{\bar \dx
    \bar P} \bar D^{\bar a \bar P} \\
  & & + 2 ||\Omega||^{-1/3} \tau_{\bar \ax \bar \bx}{}^{\bar \cx}
  (\omega_i)_{\bar \cx}{}^{\bar \dx} \bar T_{\bar \dx \bar P} C^{i \bar P} 
  + 2 ||\Omega||^{1/3} \tau_{\bar \ax \bar \bx}{}^{\bar \cx} (\eta_{a})_{\bar
    \cx}{}^{\dx} T_{\dx P} D^{a P}  \nn \\
  & & + \big[ A^1_{\bar \ax} , A^1_{\bar \bx} \big] + 2 \big[\K^\parallel_{\bar
  \ax} , A^1_{\bar \bx} \big] \; .\nn 
\end{eqnarray}
Let us compare this with the analogous formula~\eqref{F02CY} on a Calabi--Yau
manifold.  As we can see the Calabi-Yau result corresponds to the first
commutator term in the last line only, while all other terms are new.
Specifically, the first line vanishes in the Calabi-Yau case since the
covariant derivatives can be reduced to exterior derivatives which act on
closed forms. The second and third line obviously vanish for vanishing torsion
$\tau$. From Eq.~\eqref{K||} the tensor $\K^\parallel$ vanishes on a
Calabi-Yau space (in the absence of $H$-flux) and, hence, the last term also
disappears in this case. Another important remark about the above result
concerns the origin of the second line. For complex manifolds this line would
vanish identically as, in this case, the $(0,2)$ component of the field
strength can be constructed from the $(0,1)$ component of the gauge field
alone.  Then, $F_{(0,2)}$ would depend on $C$ and $D$ only but not on their
complex conjugates. This shows that our detour to real indices has been
important and, without it, we would have missed the second line of the above
result~\footnote{This can also be seen formally by observing that the torsion
  components $\tau_{\ax \bx}{}^{\bar \cx}$ are directly related to the lack of
  integrability of the almost complex structure \cite{CS}.} .

For completeness, we also present the expression for the $(1,1)$ component,
$F_{(1,1)}$, of the field strength, although this result will not be needed in
the remainder of the section.  Note that from Eq.~\eqref{tau} the first two
indices of the intrinsic torsion are both holomorphic or anti-holomorphic.
Therefore, the second and fourth line in Eq.~\eqref{Freal} do not contribute
to $F_{(1,1)}$ and we are left with
\begin{eqnarray}
  \label{hfF11}
  F_{\ax \bar \bx} & = & F^0_{\ax \bar \bx} + ||\Omega||^{-1/3}
  \nabla^{(T)}_\ax (\omega_i)_{\bar \bx}{}^{\bar 
  \cx} \bar T_{\bar \cx \bar P} C^{i \bar P} + ||\Omega||^{1/3}
  \nabla^{(T)}_\ax (\eta_a)_{\bar \bx}{}^{\cx} T_{\cx P} D^{a P} \nn \\
  && - ||\Omega||^{-1/3} \nabla^{(T)}_{\bar \bx} (\omega_i)_\ax{}^\cx T_{\cx P}
  \bar C^{i P} - ||\Omega||^{1/3} \nabla^{(T)}_{\bar \bx} (\bar \eta_{\bar
    a})_\ax{}^{\bar \cx} \bar T_{\bar \cx \bar P} \bar D^{\bar a \bar P} \\
  && + \big[A^1_\ax, A^1_{\bar \bx} \big] + \big[A^1_{\ax}, \bar K_{\bar
  \bx}^\parallel \big] + \big[\K^\parallel_{\ax} , A^1_{\bar \bx} \big] \;
  . \nn 
\end{eqnarray}
Terms proportional to $E_6$ generators in this expression can only arise from
the first commutator term in the last line, just as for Calabi-Yau manifolds.
Therefore, by virtue of Eq.~\eqref{Dterm}, the $D$-terms will be unchanged
from the Calabi-Yau case and are given by Eq.~\eqref{DCD}.

Let us now compute the $(0,3)$ component of the Chern-Simons term.  Cubic
terms in the matter fields only arise from $(A^{(1)})^3$ and should,
therefore, be unchanged from the Calabi-Yau case. This means the standard
cubic terms~\eqref{WCD} in the superpotential are also present in the
half-flat case. Terms which do not contain charged fields cancel up to higher
order terms, while linear terms are absent due to gauge invariance. Thus the
only new terms we can expect in the superpotential are quadratic terms in the
gauge matter fields. They arise from linear matter field terms in $F_{(0,2)}$,
that is the first three lines of Eq.~\eqref{F02hf} and the last term involving
$\K^\parallel$, multiplied with $A^{(1)}$, as well as from the
$\K^\parallel\wedge (A^{(1)})^2$ term contained in $A^3$.~\footnote{Note that
  quadratic terms in the charged fields which involve the ${\rm SU}(3)$ spin
  connection should vanish by the same arguments as in the Calabi--Yau case.}

Let us denote by $F^{(1)}$ the part of $F_{(0,2)}$ linear in matter fields
but excluding the contributions from $\K^\parallel$ for now.
The quadratic matter terms in the Yang-Mills  Chern-Simons form not
related to $\K^\parallel$ can then be written as 
\begin{eqnarray}
  \label{daa}
  {\rm tr} (F^{(1)} \wg A^{(1)})_{\bar \ax \bar \bx \bar \cx} & = & 6 ||\Ox||^{-2/3}
    \tau_{[\bar \ax \bar \bx}{}^ \cx (\omega_i)_{\bar \cx]}{}^{\bar \dx}
    (\omega_j)_{\cx \bar \dx} C^{i \bar P} \bar C^j{}_{\bar P} 
    + 6 ||\Ox||^{2/3} \tau_{[\bar \ax \bar \bx}{}^ \cx (\eta_a)_{\bar
      \cx]}{}^\dx (\bar \eta_{\bar b})_{\cx \dx} D^{a P} \bar D^{\bar b}{}_P
    \nn \\  
    & & + 6 \bigg[ \nabla^{(T)}_{[\bar \ax} (\eta_a)_{\bar \bx}{}^\cx
    (\omega_i)_{\bar \cx] \cx} + \tau_{[\bar \ax \bar \bx}{}^{\bar \dx}
    (\omega_i)_{\bar \cx] \cx} (\eta_a)_{\bar \dx}{}^\cx \\
    & & \qquad + \nabla^{(T)}_{[\bar \ax} (\omega_i)_{\bar \bx |\cx|} (\eta_a)_{\bar
      \cx]}{}^\cx  + \tau_{[\bar \ax \bar \bx}{}^{\bar \dx} (\eta_a)_{\bar
    \cx]}{}^\cx (\ox_i)_{\bar \dx \cx}  \bigg] C^i{}_P D^{aP} \; , \nn
\end{eqnarray}
where we have used the trace formula \eqref{tr327}.

To obtain the superpotential we have to integrate the contraction of this
formula with $\Omega^{\bar \ax \bar \bx \bar \cx}$. In the second line,
we can integrate by parts to move the derivative to $\ox_i$ and we obtain
\begin{eqnarray}
  \label{daao1}
  \int {\rm tr}(F^{(1)} \wg A^{(1)}) \wg \Omega & = & i \int \tau_{\bar \ax \bar \bx}{}^{\bar
  \dx} \bigg[ (\omega_i)_{\bar \cx \cx} (\eta_a)_{\bar \dx}{}^\cx -
  (\omega_i)_{\bar \dx \cx} (\eta_a)_{\bar \cx}{}^\cx \bigg] \Omega^{\bar \ax
  \bar \bx \bar \cx} C^i{}_P D^{aP} \nn \\ 
  & &+ 2i \int \left[\nabla^{(T)}_{\bar \ax} (\omega_i)_{\bar \bx |\cx|}
    (\eta_a)_{\bar \cx}{}^\cx  + \tau_{\bar \ax \bar \bx}{}^{\bar \dx}
    (\eta_a)_{\bar \cx}{}^\cx (\ox_i)_{\bar \dx \cx} \right] 
  \Omega^{\bar \ax \bar \bx \bar \cx} C^i{}_P D^{aP} \nn \\  
  & &+ i ||\Ox||^{-2/3} \int \tau_{\bar \ax \bar \bx}{}^ \cx (\omega_i)_{\bar
    \cx}{}^{\bar \dx} (\omega_j)_{\cx \bar \dx} \Omega^{\bar \ax \bar \bx
    \bar \cx} C^{i \bar P} \bar C^j{}_{\bar P}  \\
  & & + i ||\Ox||^{2/3} \int \tau_{\bar \ax \bar \bx}{}^ \cx (\eta_a)_{\bar 
    \cx}{}^\dx (\bar \eta_{\bar b})_{\cx \dx} \Omega^{\bar \ax \bar \bx \bar
  \cx} D^{a P} \bar D^{\bar b}{}_P \; .\nn
\end{eqnarray}
This formula looks complicated, but there are a number of simplifications.
Recall from \eqref{tau} that $\tau_{\bar \ax \bar \bx}{}^{\bar \dx}$ is a
primitive $(1,2)$ form and, therefore, the combination $\tau_{\bar \ax \bar
  \bx}{}^{\bar \dx} \Omega^{\bar \ax \bar \bx \bar \cx}$ is symmetric in the
indices $(\bar \dx, \bar \cx)$. On the other hand, the bracket in the first
line is explicitely antisymmetric in these indices. Hence, the first line
vanishes.  Further, in the second line, the indices $[\bar \ax, \bar \bx]$ are
antisymmetrised so that the covariant derivative can be converted into an
exterior derivative.  Explicitly, we have
\begin{equation}
  \label{dto}
  2 \nabla^{(T)}_{[\bar \ax} \omega_{\bar \bx] \cx} = (d \omega)_{\bar \ax \bar
    \bx \cx} - 2 \tau_{\bar \ax \bar \bx}{}^{\bar \dx} \omega_{\bar \dx \cx}
    \; ,
\end{equation}
and, therefore, the torsion drops out from the second line and only $d\ox_i$
appears. This can be replaced using the half-flat mirror
relations~\eqref{hfalg} which reduces the second line to the integral
\begin{equation}
  \label{I1}
  i e_i\int (\beta^0)_{\bar \ax \bar \bx \cx} (\eta_a)_{\bar \cx}{}^{\cx}
  \Omega^{\bar \ax \bar \bx \bar \cx} = 2 e_i K_a \; .
\end{equation}
Here $K_a$ denotes the derivative of the complex structure K\"ahler potential
$K_a = \frac{\del K(z)}{\del z^a}$.  In order to carry out the integral, we
have used the definition~\eqref{eta} of $\eta_a$  and the Kodaira
formula~\eqref{kodaira} as well as the standard choice $\cZ^0=1$.

Now we are left with having to evaluate the last two lines in Eq.~
\eqref{daao1}. First note from \eqref{tau} that the contraction of the torsion
$\tau_{\bar \ax \bar \bx}{}^\cx $ with $\Omega^{\bar \ax \bar \bx \bar \cx}$
can be written as
\begin{equation}
  \label{hfid}
  \tau_{\bar \ax \bar \bx}{}^{\cx} \Omega^{\bar \ax \bar \bx \bar \cx} =
  \frac{i}{4 \cK} e_i g^{ij} (\omega_j)^{\cx \bar \cx} \; .
\end{equation}
Then, the last two lines in Eq.~ \eqref{daao1} lead to the integrals
\begin{equation}
  \label{sdef}
  \sigma_{lij} = \frac{i}{4 \cK} \int (\omega_l)^{\cx \bar \cx}
  (\omega_i)_{\bar \cx}{}^{\bar \dx} (\omega_j)_{\cx \bar \dx} \; ,
\end{equation}
and 
\begin{equation}
  \label{tsdef}
  \tilde \sigma_{j a \bar b} = \frac{i}{4 \cK} \int (\omega_j)^{\cx \bar \cx}
  (\eta_a)_{\bar \cx}{}^\dx (\bar \eta_{\bar b})_{\cx \dx} \; .
\end{equation}
These integrals are non-standard on a Calabi--Yau manifold and
presumably difficult to compute. However, we will not need their general values
but merely their contractions with the K\"ahler moduli $v^i$.
Using \eqref{Kms}, \eqref{g11} and \eqref{g21eta} these contractions can be
explicitly computed and we find
\begin{equation}
  \label{vsig}
  v^l \sigma_{lij} = g_{ij} \; ,\qquad v^j \tilde \sigma_{ja \bar b} = \frac14 g_{a \bar b} \; ,
\end{equation}
where $g_{ij}$ and $g_{a \bar b}$ are the K\"ahler and complex structure moduli
space metrics.

Combining these results we can finally write Eq.~\eqref{daao1} as
\begin{equation}
  \label{F1A1}
  \int (F^{(1)} \wedge A^{(1)}) \wedge \Omega = \alpha' \bigg[ i e_k g^{kl}
    \left(||\Omega||^{-2/3} \sigma_{lij} C^{i \bar P} \bar C^j{}_{\bar P} +
    ||\Omega||^{2/3} \tilde \sigma_{la \bar b} D^{aP} \bar D^{\bar b}{}_P
    \right) + 2 e_i K_a C^{i \bar P} D^a{}_{\bar P} 
    \bigg] \; .
\end{equation}
To obtain all quadratic matter field terms in the gravitino mass we still have
to add the terms related to $\K^\parallel$. They arise from terms of the form
$\K^\parallel\wedge (A^{(1)})^2$ which are contained in both the $F\wedge A$
and $A^3$ terms of the Yang-Mills Chern-Simons form.
Taking all the factors into account we get the following contribution to the
gravitino mass from terms proportional to $\K^\parallel$ 
\begin{eqnarray}
  \label{W8}
  \left . \int \omega_{YM} \wedge \Omega \right|_{\K} & = & -2 i \ap
  \int \K^\parallel_{\bar \ax \ax \bar \dx} (\omega_i)_{\bar \bx}{}^{\bar \dx}
  (\eta_a)_{\bar \cx}{}^\ax \Omega^{\bar \ax \bar \bx \bar \cx} C^i_P
  D^{aP} \\
  & = & i \ap e_j t^j  \int (\beta^0)_{\bar \ax \ax \bar \dx}
  (\omega_i)_{\bar \bx}{}^{\bar \dx} (\eta_a)_{\bar \cx}{}^\ax
  \Omega^{\bar \ax \bar \bx \bar \cx} C^i_P D^{aP} \; , \nn
\end{eqnarray}
where we have used Eq.~\eqref{K||}.
This integral is similar to the ones discussed above and can, in fact, be
written in terms of $\tilde \sigma_{i a \bar b}$ defined in \eqref{tsdef}.
However, we will not need its explicit form and simply introduce the symbol
$\Sigma_{ia}$ 
\begin{equation}
  \label{Sigma}
  \Sigma_{ia} = i \int (\bx^0)_{\bar \ax \ax \bar \dx} (\omega_i)_{\bar
  \bx}{}^{\bar \dx} (\eta_a)_{\bar \cx}{}^{\ax} \Omega^{\bar \ax \bar \bx \bar
  \cx} \; ,
\end{equation}
such that 
\begin{equation}
  \label{W8fin}
   \left . \int \omega_{YM} \wedge \Omega \right|_{\K} = \ap e_j t^j
   \Sigma_{ia} C^i_P D^{aP} \; .
\end{equation}

To summarise this section let us write the final formula for the gravitino
mass at order $\ap$ which is given by the sum of Eqs.~\eqref{WCD},
\eqref{Whf}, \eqref{F1A1} and \eqref{W8fin} and reads
\begin{eqnarray}
  \label{m32hf}
  m_{3/2}\equiv e^{K/2} W & = & e^{K_0/2} \Bigg\{e_i t^i -\frac13 \ap \left[\bar j_{\bar P \bar R \bar S}
    \cK_{ijk} C^{i \bar P} C^{j \bar R} C^{k \bar S} + j_{PRS} \tilde \cK_{abc}
    D^{a P} D^{bR} D^{cS} \right] \\ 
  & & + \alpha' \bigg[ i e_k g^{kl} \left(||\Omega||^{-2/3} \sigma_{lij} C^{i
      \bar P} \bar C^j{}_{\bar P} + ||\Omega||^{2/3} \tilde \sigma_{la \bar b}
    D^{aP} \bar D^{\bar b}{}_P \right) + 2 e_i K_a C^{i \bar P} D^a{}_{\bar P}
    \bigg] \nn \\ 
  & & + \ap e_i t^i \Sigma_{ja} C^j_P D^{aP} \Bigg\} \; . \nn
\end{eqnarray}
Here, $K_0$ stands for the K\"ahler potential at zeroth order in $\ap$.
This is the one of the main results of the paper. In the following sections we
will analyse its interpretation and implications for the four-dimensional
effective theory.

\subsection{Four-dimensional effective theory}
\label{sec:KW}

Eq.~\eqref{m32hf} provides us with the with the supergravity $G$-function,
$G=K+\ln |W|^2=\ln |m_{3/2}|^2$. This, together with the gauge
kinetic function which has already been computed in Ref.~\cite{GLM},
completely determines the four-dimensional supergravity Lagrangian. It seems,
we should, therefore, be able to find the complete low-energy theory from the
results so far. However, Eq.~\eqref{m32hf} as stands is still expressed in
terms of the 10-dimensional fields and it first needs to be re-written in
terms of the correct four-dimensional superfields. In other words, we need to
know the definition of the four-dimensional superfields in terms of the
underlying 10-dimensional fields. It is not obvious that this information can
be extracted from the above results.  To analyse the situation it is useful to
compare Eq.~\eqref{m32hf} with a general expression for the gravitino mass,
expanded to linear order in $\ap$.  Let us denote by $\phi_0$, $K_0$ and $W_0$
the moduli fields, the K\"ahler potential and the superpotential at zeroth order
in $\ap$ and by $\phi$, $K$ and $W$ their counterparts at order first order in
$\ap$. We can write
\begin{eqnarray}
  \phi &=& \phi_0+\ap \delta\phi \; , \nn \\
  K(\phi )&=&K_0(\phi_0 )+\ap \delta K \; , \label{Kgen} \\
  W(\phi ) &=& W_0(\phi_0)+\ap (\delta W+\partial_\phi W \delta\phi )\; , \nn
\end{eqnarray}
where $\delta\phi$ is a correction to the definition of the moduli fields
which is expected~\cite{witten} at order $\ap$. Further $\delta K$ and $\delta
W$ are the changes of the K\"ahler potential and superpotential~\footnote{It is
  convenient to separate out the change of the superpotential due to the
  re-definition of the moduli fields explicitly while writing the change in
  the K\"ahler potential as a single term.}  at order $\ap$. The gravitino mass
can then be expanded as~\footnote{Since we are working to first order in
  $\ap$, we do not need to distinguish between corrected and uncorrected
  quantities in the order $\ap$ part of this expression.}
\begin{equation}
  \label{KWfin}
  m_{3/2}=e^{K/2} W = e^{K_0(\phi_0)/2} \left[W_0(\phi_0) + \ap\left( \dx W +
   W \dx K  + \del_\phi W \dx \phi \right)\right] \; .
\end{equation}
Let us now compare this general expression to our explicit result for the
gravitino mass~\eqref{m32hf}.  Clearly, the first term in Eq.~\eqref{m32hf}
corresponds to the superpotential at zeroth order in $\ap$, that is, to the
term $W_0(\phi_0)$ in our general notation. The rest of the first line is the
well-known cubic superpotential for the matter fields which arises at order
$\ap$ and it should be part of $\delta W$. All other terms in
Eq.~\eqref{m32hf} are of order $\ap$ and non-holomorphic and, hence, must
correspond to the last two terms in Eq.~\eqref{KWfin}, that is, they must be
due to $\ap$ corrections to the definition of the moduli fields or to the
K\"ahler potential. Given that we have a K\"ahler moduli superpotential at zeroth
order in $\ap$ we indeed need correction terms which convert this
superpotential into a function of the proper order $\ap$ superfields. What we
have to decide is which of the terms in the second and third line of
\eqref{m32hf} are absorbed into a re-definition of the K\"ahler moduli $t^i$. It
turns out the correct choice is to absorb all terms in the second line of
\eqref{m32hf} into $t^i$ while interpreting the term in the third line as a
correction to the K\"ahler potential. This is suggested by the fact that
corrections to $K$ should appear multiplied with $W$ (as in Eq.~\eqref{KWfin})
which is only the case for the last term in \eqref{m32hf}. Also, we know from
the standard Calabi-Yau case~\cite{witten} that the first two terms in the
second line should definitely be part of the re-definition of $t^i$, so the
only non-trivial question is really about the last two terms in \eqref{m32hf}.
We will check towards the end of this section that our choice for these
remaining two terms is indeed correct.

Let us now formalise the previous discussion. We write the relation between
the zeroth order K\"ahler moduli $t^i$ and their order $\ap$ counterparts $T^i$
as
\begin{equation}
  \label{corrT}
  T^i = t^i + \alpha' Y^i \; .
\end{equation}
where the correction terms $Y^i$ are explicitly given by
\begin{equation}
  \label{T1}
  Y^i = i g^{ij} \big(||\Omega||^{-2/3}\sigma_{jkl} C^{k
    \bar P} \bar C^l{}_{\bar P}+ ||\Omega||^{2/3}\tilde \sigma_{ja \bar b}
    D^{aP} \bar D^{\bar b}{}_P \big) +2 K_a C^{i \bar P} D^a{}_{\bar P}\; .
\end{equation}
From Eq.~\eqref{m32hf}, the superpotential is then given by
\begin{equation}
  \label{Wfin}
  W = e_i T^i - \frac{\ap}{3} \bar j_{\bar P \bar R \bar S} \cK_{ijk} C^{i
    \bar P} C^{j \bar R} C^{k \bar S} - \frac{\ap}{3} j_{PRS} \tilde \cK_{abc}
    D^{a P} D^{bR} D^{cS} \; .
\end{equation}
Note that the torsion part of the superpotential has absorbed the terms in the
second line of \eqref{m32hf} and, as a result, is now expressed in terms of
the corrected superfields, $T^i$, as is should.

The only remaining term is the last one in \eqref{m32hf}. It gives rise to a
K\"ahler potential correction so that the total K\"ahler potential,
$K=K_0(\phi_0)+\delta K$, can be written as
\begin{equation}
  \label{K1}
  K=K_0(s,\bar{s},t,\bar{t},z,\bar{z})+\alpha' \Big[ \Sigma_{ia} C^i_P D^{aP}
  +{\rm c.c.}\Big]\; , 
\end{equation}
where the moduli K\"ahler potential $K_0$ is given in Eq.~\eqref{K0}. The K\"ahler
moduli part, $K_K$, of $K_0$ still needs to be expressed in terms of the
corrected moduli fields $T^i$, so we write
\begin{equation}
  \label{Kty}
  K_K(t, \bar t) = K_K(T-\ap Y, \bar T- \ap \bar Y) = K_K(T, \bar T) - \ap K_i
  Y^i - \ap \bar K_i \bar Y^i + \mathcal{O}(\ap^2) \; ,
\end{equation}
where $K_i$ is the derivative of the K\"ahler potential with respect to
$t^i$. Using the Calabi--Yau identity
\begin{equation}
  \label{CYid}
  K_i g^{ij} = 2i v^j \; ,
\end{equation}
together with Eqs.~\eqref{vsig} we find 
\begin{equation}
  \label{KY}
  K_i Y^i = - 2 ||\Omega||^{-2/3} g_{kl} C^{k \bar P}
    \bar C^l{}_{\bar P}-\frac{1}{2} ||\Omega||^{2/3}g_{a \bar b} D^{aP}
    \bar D^{\bar b}{}_P  +2 K_i K_a C^{i \bar P} D^a{}_{\bar P}\; .
\end{equation}
The K\"ahler potential~\eqref{K1} can then be written as 
\begin{eqnarray}
K&=&K_0(s,\bar{s},T,\bar{T},z,\bar{z})+\alpha' \left[
  4 ||\Omega||^{-2/3} g_{ij} C^{i \bar P} \bar C^j{}_{\bar P} +
  ||\Omega||^{2/3} g_{a \bar b} D^{a P} \bar D^{\bar b}{}_P \right.\nn\\
  &&\hskip 4cm\left.+\big((\Sigma_{ia}-2K_i K_a) C^i{}_P D^{aP} + {\rm c.c.} \big) \right]\label{K11} \; .
\end{eqnarray}
Having corrected the K\"ahler moduli at order $\ap$ it seems likely the same has
to be done to the complex structure moduli, so we write the $\ap$ corrected
complex structure moduli $Z^a$ as
\begin{equation}
  Z^a=z^a+\ap Y^a\; .
\end{equation}
From the result~\eqref{m32hf} we have no direct information about the
corrections $Y^a$ since the zeroth order superpotential is independent of the
complex structure moduli. One guess might be that the additional $\delta K$
term (the term proportional to $\Sigma_{ia}$ in Eq.~\eqref{K11}) we have found
is responsible for the re-definition of the complex structure moduli. This
would imply that
\begin{equation}
  Y^a={\Sigma_{ia}}^bC^i{}_P D^{aP} \label{Ya}
\end{equation}
for some tensor ${\Sigma_{ia}}^b$ with the property
$K_b{\Sigma_{ia}}^b=\Sigma_{ia}$. Later we will introduce $H$-flux which
provides us with a zeroth order superpotential for the complex structure
moduli and explicit information about the redefinition of $z^a$. We will then
confirm the above expression for $Y^a$. Accepting our guess for now we can
write the K\"ahler potential as
\begin{eqnarray}
  K&=&K_0(s,\bar{s},T,\bar{T},Z,\bar{Z})+\alpha' \left[
    4 ||\Omega||^{-2/3} g_{ij} C^{i \bar P} \bar C^j{}_{\bar P} +
    ||\Omega||^{2/3} g_{a \bar b} D^{a P} \bar D^{\bar b}{}_P \right.\nn\\
  &&\hskip 4cm\left.-2\big(K_i K_a C^i{}_P D^{aP} + {\rm c.c.} \big)
  \right]\label{Kfin} \; . 
\end{eqnarray}
Eqs.~\eqref{Kfin} and \eqref{Wfin} represent our final result for the order
$\ap$ K\"ahler potential and superpotential from standard embedding
compactifications of the heterotic string on half-flat mirror manifolds. The
matter field part is $E_6$ invariant and identical to the one found for
Calabi-Yau compactifications with standard embedding. The only difference to
the Calabi-Yau case is the zeroth order torsion superpotential for the K\"ahler
moduli which is simply added to the standard cubic superpotential for the
matter fields.  Although the $CD$ terms in the K\"ahler potential~\eqref{Kfin}
(and the field re-definitions~\eqref{T1}) look unconventional they are
independent of the torsion parameters $e_i$ and should, therefore, be already
present in the Calabi-Yau case. To our knowledge they have not been explicitly
computed before, although their possible existence has been anticipated in
Ref.~\cite{DKL}. These terms do not contribute to the matter field kinetic
terms (although they do contribute to mixed matter field/moduli kinetic terms)
and, in the Calabi-Yau case, they affect the scalar potential only at higher
order in $\ap$. It is not surprising, therefore, that these terms are usually
omitted.  A curious feature of our result is that the order $\ap$
re-definition of K\"ahler and complex structure moduli is quite different, see
Eqs.~\eqref{T1} and \eqref{Ya}. In particular, $C\bar{C}$ and $D\bar{D}$ terms
appear for the K\"ahler moduli only. This means that the standard kinetic terms
for both the $(1,1)$ and $(2,1)$ matter fields are linked to the re-definition
of the $(1,1)$ moduli.

Should we have expected new terms in the matter field sector compared to the
Calabi-Yau case? Given that our set-up leads to an $E_6$ invariant low-energy
theory, the only additional terms allowed from gauge invariance are
${C^i}_PD^{aP}$ terms in the superpotential. We know that such terms are
definitely absent in the Calabi-Yau case and this can be understood from the
fact that we have required that the matter fields $C$ and $D$ in
Eq.~\eqref{chfdef1} be massless. By turning on fluxes we may expect that some
of these fields become massive, but the above calculation shows that a
supersymmetric mass term is not generated.\footnote{It is not hard to see that
  at the level of the $N=1$ potential, terms of the type $(e_i T^i) g_{jk}
  C^j_P \bar C^{kP}$ and similar ones for the $D$-fields are in fact generated
  at first order in $\ap$ making these fields indeed massive. Note that this
  is only possible at this order in $\ap$ due to the appearance of the zeroth
  order superpotential \eqref{Whf}.} We
interpret this as an indication that 2-index couplings (fluxes),
$\lambda_{ia}$, are needed for these terms to appear in the superpotential. We
will in fact see that for the generalised half-flat manifolds discussed in
Section~\ref{sec:genhf}, for which torsion parameters have one K\"ahler and
one complex structure index, $CD$ superpotential terms indeed arise.

One obvious simple check of our results is to compare the D-term, as obtained
from the Gukov-type formula~\eqref{Dterm}, with the four-dimensional
supergravity expression \eqref{Dsugra}
after inserting the above results for K\"ahler potential and superpotential.
 We recall that the Gukov-type
formula predicts the D-terms for half-flat mirror manifolds should be
unchanged from the Calabi-Yau case. This is indeed what one finds when
inserting ~\eqref{Kfin} and \eqref{Wfin} into the supergravity
formula~\eqref{Dsugra}.

\subsection{Including H-flux}
\label{sec:Hflux}

An obvious extension of our set-up is to include NS-NS flux. This will
generate a zeroth order superpotential for the complex structure moduli, in
addition to the K\"ahler moduli superpotential from torsion already present. As
indicated before, this can provide us with additional information about the
complex structure moduli and consistency checks of our results.

For simplicity, we start with NS-NS flux of the form $H_{\rm flux} = p_a \bx^a$, with flux parameters $p_a$.
This leads to a zeroth order superpotential contribution
\begin{equation}
  \label{WH}
  W_{0,{\rm flux}} = \int \Omega \wedge H = p_a z^a \; .
\end{equation}
which, in analogy with the torsion superpotential, is linear in the moduli.

Does the NS-NS flux lead to any corrections at first order in $\ap$? We have
to remember that, via Eq.~\eqref{modsc}, $H$ appears in the connection which
enters the Bianchi identity. The resulting change in the gravitino mass can be
computed directly from Eq.~\eqref{W8} with $\K$ replaced by $H_{\rm flux}$ and
making use of formulae \eqref{bchi}.
Together with the zeroth order term~\eqref{WH} this leads to the following
additional terms in the gravitino mass due to flux
\begin{equation}
  \left. e^{K/2}W\right|_{\rm flux}=e^{K_0/2}\left[p_az^a
    +\ap (p_b+p_cz^cK_b){\Sigma_{ia}}^b C^i_P D^{aP} \right]\; ,
  \label{m32flux}
\end{equation}
where we have defined the quantity
\begin{equation}
  {\Sigma_{ia}}^b = - \frac{i}{\int \Omega \wedge \bar \Omega}
  g^{\bar{a}b}\int (\bar \chi_{\bar a})_{\bar \ax \ax \bar \dx} 
  (\omega_i)_{\bar \bx}{}^{\bar \dx} (\eta_a)_{\bar \cx}{}^\ax \Omega^{\bar
    \ax \bar \bx \bar \cx} \; .
\end{equation}
Let us now analyse this result by comparing it to the general
expression~\eqref{KWfin} for the gravitino mass, as we did before. The order
$\ap$ terms in Eq.~\eqref{m32flux} are non-holomorphic and, hence, they should
correspond to either corrections to the K\"ahler potential or re-definitions of
moduli fields.  The last term in Eq.~\eqref{m32flux} is proportional to the
flux superpotential and this suggests it should be viewed as a correction to
the K\"ahler potential. This interpretation is, in fact, required for
consistency, given that we have declared the last term in \eqref{m32hf} to be
a K\"ahler potential correction as well. Indeed, in the presence of
flux, we need another term to combine with the last one in Eq.~\eqref{m32hf}
to produce the total torsion and flux superpotential as a pre-factor. One
can check, using relation \eqref{bchi}, that 
\begin{equation}
  K_b{\Sigma_{ia}}^b=\Sigma_{ia}\; ,
\end{equation}
with $\Sigma_{ia}$ defined in Eq.~\eqref{Sigma} which provides a confirmation of
this interpretation.
The second term in Eq.~\eqref{m32flux}, on the other hand, coincides with the
correction~\eqref{Ya} to the definition of the complex structure moduli which
we have anticipated earlier. Hence, this term combines with the flux
superpotential and changes the moduli $z^a$ into their $\ap$ corrected
counterparts $Z^a$.

To summarise, we have confirmed our earlier result~\eqref{Kfin} for the K\"ahler
potential and the superpotential is given by~\eqref{Wfin} plus the addition
flux contribution
\begin{equation}
W_{\rm flux}=p_aZ^a\;.
\end{equation}

\subsection{Consistency with compactification results}

The identification of low-energy data from the gravitino mass~\eqref{m32hf},
\eqref{m32flux} in the last sub-section has, in parts, relied on a suggestive
interpretation rather than strict conclusion. It would, therefore, be
desirable to have an independent and meaningful check through a direct
compactification of the 10-dimensional theory. We have argued before that this
is a difficult task, firstly due to the large number of terms in the potential
and secondly due to the presence of certain integrals which are not standard
on Calabi--Yau manifolds. We have already encountered such integrals in the
calculation of the gravitino mass, although we have managed to proceed without
knowing their explicit form.  One way to simplify the calculation is to consider
an underlying Calabi--Yau manifold with only a single K\"ahler modulus. Then
the forms $\ox_i$ can be replaced by the almost complex structure $J$ and the
integrals involved become significantly simpler. The number of terms is also
reduced, although it is still considerable.  In addition to assuming
$h^{1,1}=1$, we will, therefore, focus on a specific class of terms, namely
scalar potential terms of the form $DD \bar C$ and $CC \bar D$ together with
their complex conjugates.

Let us explain how these terms appear when compactifying the 10-dimensional
action. First of all, cubic terms in the matter fields appear from $F^2$,
taking terms linear in the charged fields in one $F$ (either from the explicit
terms in \eqref{Freal} or from the last commutator) and terms quadratic in the
matter fields in the second $F$ (from the first commutator in \eqref{Freal}).
Another source for cubic matter field terms is $H^2$ with one $H$ taken to be
the zeroth order part~\eqref{Hhf} from torsion and the other the Chern-Simons
form (in fact only the term $A^3$ can contribute). For the case of one K\"ahler
modulus, it is not hard to check that all of the derivative terms from
\eqref{Freal} drop out -- either directly or after integration by parts -- and
only the linear terms with no derivative contribute. After a long but
straightforward calculation one obtains
\begin{eqnarray}
  \label{VCCD}
  V & = & \ldots + \ap \frac{e^{-2 \phi_4}}{\cK} \left[ \frac{1}{\cK
      ||\Omega||^{8/3}} \left(\frac{3i}2 \; \frac{eb}{v} + \frac{e}2 \right)
    K_{\bar c} g^{\bar c c} \tilde \cK_{abc} j_{PRS} D^{aP} D^{bR} \bar C^S
  \right. \nn \\ 
  & & \left. + \frac{v^2}{3 \cK ||\Omega||^{4/3}} \left(\frac{3i}2 \;
  \frac{eb}{v} + \frac{e}2 \right) K_{\bar a} j_{\bar P \bar R \bar S} C^{\bar
  P} C^{\bar R} \bar D^{\bar a \bar S} \right]  + {\rm c.c.} \; ,
\end{eqnarray}
where the dots stand for other type of terms we have not computed here. 

On the supergravity side, starting from the K\"ahler potential \eqref{Kfin}
and superpotential \eqref{Wfin}, we can compute the above terms for general
$h^{1,1}$ and only at the very end take the limit of one K\"ahler modulus. We
note that the $|W|^2$ term in the supergravity potential cannot produce the
type of terms considered here.  Therefore we only have to consider the
$F$-part of the supergravity potential.  To get a contribution proportional to $C^2$ or $D^2$
to an F-term we need to consider the derivatives $\del_C W$ or $\del_D W$,
respectively. These derivatives are already at order $\ap$ and, hence, need to
to be multiplied by an order zero piece. It is clear that the $C\bar{C}$ and
$D\bar{D}$ terms in the matter field K\"ahler potential cannot lead to cubic
terms mixing $C$ and $D$.  Therefore the term $CD$ in the K\"ahler potential
is crucial in order to reproduce the terms in \eqref{VCCD}. To see precisely
which terms do contribute one has to compute the K\"ahler metric, including the
complex structure moduli, from \eqref{Kfin} and invert this metric at order
$\ap$.  Schematically, one then finds the following relevant terms
\begin{equation}
  V \sim D_CW \left(g^{\bar C C} D_{\bar C} \bar W + g^{C \bar z} D_{\bar z}
  \bar W + g^{C \bar T} D_T \bar W \right) \; ,
\end{equation}
plus similar terms with $C$ and $D$ exchanged. From the first term in the
bracket one keeps the derivative on the K\"ahler potential times $W_0$, from
the second term the derivative on the complex structure K\"ahler potential
times $W_0$ while the full $D_T \bar W_0$ contributes from the last term.
Computing explicitely all terms and taking the limit of one K\"ahler modulus,
it is not hard to see that the result indeed reproduces \eqref{VCCD}. This
constitutes a powerful check of our results. In particular, it confirms that
the $CD$ terms should indeed be present in the K\"ahler potential
\eqref{Kfin} and that the identifications of various terms in \eqref{m32hf}
was correct.

\section{Generalised half-flat manifolds}
\label{sec:genhf}

In the final part of this paper let us discuss an extension of the results
obtained in the previous section to more general manifolds with ${\rm SU}(3)$
structure which we refer to as generalised half-flat manifolds. These
manifolds were proposed in Ref.~\cite{ferrara,GLW}.  Working out the effective
theories associated to these manifolds is not conceptually new, but rather a
straightforward generalisation of the results obtained in the previous
section.

Let us briefly present the setup for these compactifications, relying on the
conventions of Ref.~\cite{dCGLM}. We consider a manifold with ${\rm SU}(3)$
structure and two-forms $\omega_i$ and three-forms $(\ax_A,\bx^B)$ which obey
the following algebra
\begin{eqnarray}
  \label{exthf}
  d \omega_i & = & p_{iA} \bx^A - q_i^A \ax_A \; ,\nn \\
  d \ax_A  & = & p_{iA} \tox^i \; , \qquad d \bx^A = q_i^A \tox^i \; ,\\
  d \tox^i & = &0 \; , \nn
\end{eqnarray}
with the constants $p_{iA}$ and $q_i^A$ subject to the constraints
\begin{equation}
  \label{constraint}
  p_{iA} q_j^A - p_{jA} q_i^A =0 \; .
\end{equation}
These relations replace the analogous relations \eqref{hfalg} for half-flat
mirror manifolds.  In addition, it is assumed that the link of these manifolds
with underlying Calabi-Yau manifolds is precisely as for half-flat mirror
manifolds. Half-flat mirror manifolds corresponds to the particular choice
$p_{i0}=e_i$, $p_{ia}=0$ and $q_i^A=0$.

The zeroth order superpotential for generalised half-flat manifolds is more
complex than the one for half-flat mirror manifolds, Eq.~\eqref{Whf}, and contains
mixed K\"ahler and complex moduli terms as well. It is given by~\cite{dCGLM}
\begin{equation}
  \label{Wext0}
  W_0 = p_{iA} t^i \cZ^A - q_i^A t^i \cG_A \equiv E_i t^i \, ,
\end{equation}
where $\cG_A$ are the derivatives of the complex structure
prepotential $\cG$. To make the analysis similar to the half-flat case we have
introduced the notation 
\begin{equation}
  \label{Ei}
  E_i(\cZ) = p_{iA} \cZ^A - q_i^A \cG_A\; .
\end{equation}
With this notation, much of the calculations for half-flat mirror manifolds
can be carried over to the present case by replacing the torsion parameters
$e_i$ with $E_i$.  In particular, Eq.~\eqref{hfid} still holds with this
replacement. Hence, we can directly obtain the result for the last integrals
in Eq.~\eqref{daao1}. Working out the second line in Eq.~\eqref{daao1} however, 
requires some modifications. The result previously given by Eq.~\eqref{I1} now
changes to
\begin{equation}
  \label{I1g}
  i \int (d \omega_i)_{\bar \ax \bar \bx \cx} (\eta_a)_{\bar
  \cx}{}^{\cx} \Omega^{\bar \ax \bar \bx \bar \cx} = - 2 \int d \omega_i
  \wedge \chi_a = 2(p_{ia} - q_i^B \cG_{Ba}) + 2K_a E_i 
\end{equation}
where the last equality follows from straightforward computations using
the standard Calabi-Yau relations \eqref{bchi}. Finally one can
show that $\K^\parallel$ in 
Eq.~\eqref{K||} can now be written as\footnote{It is not hard to check that
  for the generalised half-flat case the the tensor $\K^\perp$ takes a form
  similar to the one in the half-flat mirror case, Eq.~\eqref{K|-}. This
  allows us again to absorb these pieces into a redefinition of the charged
  fields $C$ and $D$.} 
\begin{equation}
  \label{K||ghf}
  (\K^\parallel_{\bar \ax})_{\ax \bar \bx} = - \frac12 t^i d \ox_i = - \frac12
  t^i( p_{iA} \bx ^A - q_i^A \ax_A) \; .
\end{equation}
With this, and using again formulae \eqref{bchi}, the result in
Eq.~\eqref{W8fin} becomes 
\begin{equation}
  \left . \int \Omega_{YM} \wedge \Omega \right|_\K = \ap t^i  \left[(p_{ia} -
  q_i^A \cG_{Aa}) + K_a (p_{ia} \cZ^A - q_i^A \cG_A) \right ] \Sigma_{jb}{}^a
  C^j_P D^{bP} \; .
\end{equation}
Collecting the above contributions, the final formula for the gravitino mass
in this more general case then takes the form
\begin{eqnarray}
  \label{m32ghf}
  e^{K/2} W & = & e^{K_0/2} \Bigg\{E_i t^i -\frac13 \ap
    \left[\bar j_{\bar P \bar R \bar S} \cK_{ijk} C^{i \bar P} C^{j \bar R}
    C^{k \bar S} + j_{PRS} \tilde \cK_{abc} D^{a P} D^{bR} D^{cS} \right] \\ 
  & & + \alpha' \bigg[ i E_k g^{kl} \left(||\Omega||^{-2/3} \sigma_{lij} C^{i
      \bar P} \bar C^j{}_{\bar P} + ||\Omega||^{2/3} \tilde \sigma_{la \bar b}
    D^{aP} \bar D^{\bar b}{}_P \right) + 2 E_i K_a C^{i \bar P} D^a{}_{\bar P}
    \bigg] \nn \\ 
  & & + \ap \left[(p_{ia} -  q_i^A \cG_{Aa})t^i + K_a E_i t^i \right ]
    \Sigma_{jb}{}^a C^j_P D^{bP} + 2 \ap (p_ia - q_i^B \cG_{Ba}) C^i_P D^{aP}
    \Bigg\} \; . \nn 
\end{eqnarray}
Given the expression~\eqref{Wext0} for the torsion superpotential,
it follows that the interpretation of most terms above is the same as for the
half-flat mirror case \eqref{m32hf}: the first line is part of the
superpotential at order $\ap$, the second line should be interpreted as a
redefinition of the K\"ahler moduli $t^i$ like in Eqs.~\eqref{corrT} and
\eqref{T1} while the first terms in the last line are analogous to the last
terms in Eq.~\eqref{m32flux} and correspond to the redefinition of the complex
structure moduli \eqref{Ya} and the change in the K\"ahler potential from
Eq.~\eqref{K11}. The only difference compared to the cases
studied before is the last term above. As it is holomorphic
there is no need to absorb it into a redefinition of moduli and it turns to be
part of the superpotential. On general grounds, this is actually not
surprising given that we now have the couplings $p_{iA}$ and $q_i^A$ which
allow holomorphic (and gauge invariant) terms such as $p_{ia} C^i_P D^{aP}$.

To summarise, for the generalised half-flat manifolds, characterised by the
algebra~\eqref{exthf}, the K\"ahler potential is unchanged from the half-flat
mirror case and still given by Eq.~\eqref{Kfin}, while the superpotential now
reads
\begin{eqnarray}
  \label{Wfine}
  W & = & p_{i0} T^i + p_{ia} T^i Z^a - q_i^A T^i \cG_A(Z) 
  + 2 \alpha' (p_{ia} - q_i^A \cG_{Aa}) C^i{}_P D^{aP} \\[2mm]
  & & - \frac{\alpha'}{3} \left[ \cK_{ijk} j_{\bar P \bar R \bar S} C^{i \bar
  P} C^{j \bar R} C^{k \bar S} + \tilde \cK_{abc} j_{PRS} D^{aP} D^{bR} D^{cS}
  \right] \; .\nn
\end{eqnarray}

\section{Conclusions}
\label{sec:con}

In this paper, we have studied heterotic string compactifications at order
$\ap$ on specific classes of manifolds with $SU(3)$ structure, namely
half-flat mirror manifolds and their generalisations.  These manifolds are
related to underlying Calabi-Yau manifolds which facilitates the explicit
computation of the associated four-dimensional effective theories. In order to
solve the Bianchi identity, we have employed the simplest possibility for the
choice of the internal gauge bundle, a variant of the well-known standard
embedding. The spin connection of half-flat manifolds has in general ${\rm
  SO}(6)$ holonomy which suggests a low-energy gauge group ${\rm SO}(10)$.
However, we were able to absorb the pieces of the connection which would have
been responsible for this breaking to ${\rm SO}(10)$ into a vacuum
redefinition of the matter fields. For the fields re-defined in this way, we
found that the $E_6$ gauge symmetry is restored. The four-dimensional
effective theory contains a dilaton $s$, $h^{1,1}$ K\"ahler moduli $T^i$ and
$h^{2,1}$ complex structure moduli $Z^a$, where $h^{1,1}$ and $h^{2,1}$ are
the Hodge numbers of the associated Calabi-Yau manifolds. In addition, there
are $h^{1,1}$ matter fields $C^{i\bar{P}}$ in the ${\bf \overline{27}}$
representation and $h^{2,1}$ matter fields $D^{aP}$ in the ${\bf 27}$
representation of $E_6$. Hence, the low-energy field content is precisely the
same as for analogous Calabi-Yau compactifications of the heterotic string.
The half-flat manifolds are defined in the large radius and large complex
structure limit and it is in this limit that the effective four-dimensional
theory has been derived. For the K\"ahler potential we find
\begin{eqnarray}
  K&=&K_s(s,\bar{s})+K_K(T,\bar{T})+K_{cs}(Z,\bar{Z})\nn\\
  &&+\alpha' \left[
    4 ||\Omega||^{-2/3} g_{ij} C^{i \bar P} \bar C^j{}_{\bar P} +
    ||\Omega||^{2/3} g_{a \bar b} D^{a P} \bar D^{\bar b}{}_P 
    -2\big(K_i K_a C^i{}_P D^{aP} + {\rm c.c.} \big) \right]\label{Kfin1} \; ,
\end{eqnarray}
where $||\Omega||^2=\exp (K_K-K_{cs})$ and $K_s$, $K_K$ and $K_{cs}$ are the
standard Calabi-Yau K\"ahler potentials for the dilaton, the K\"ahler moduli
and the complex structure moduli. Explicit formulae for these K\"ahler
potentials are given in Eqs.~\eqref{Ks}, \eqref{Kt} and \eqref{Kcs1},
respectively. The superpotential is given by
\begin{eqnarray}
  W & = & p_{i0} T^i + p_{ia} T^iZ^a - q_i^A T^i \cG_A(Z) 
  + 2 \alpha' (p_{ia} - q_i^A \cG_{Aa}) C^i{}_P D^{aP} \\[2mm]
  & & - \frac{\alpha'}{3} \left[ \cK_{ijk} j_{\bar P \bar R \bar S} C^{i \bar
  P} C^{j \bar R} C^{k \bar S} + \tilde \cK_{abc} j_{PRS} D^{aP} D^{bR} D^{cS}
  \right] \; .\nn
\end{eqnarray}
where $\cG_A$ are the derivatives of the complex structure pre-potential
${\cal G}$, given in Eq.~\eqref{csprep}, and $\kappa_{ijk}$ and
$\tilde\kappa_{abc}$ are the intersection numbers of the underlying Calabi-Yau
manifolds and its mirror. Further $j_{PRS}$ and $j_{\bar{P}\bar{R}\bar{S}}$
are the cubic $E_6$ invariant tensors. This expression for the superpotential
is given for the generalised half-flat manifolds discussion in
Section~\ref{sec:genhf}. To specialise to half-flat mirror manifolds one
should set $p_{i0}=e_i$, $p_{ia}=0$ and $q_i^A=0$. Note that in this case the
$CD$ mass term in $W$ vanishes.  NS-NS flux with flux parameters $\epsilon_A$
and $\mu^A$ leads to an additional superpotential of the usual form
\begin{equation}
  W_{\rm flux}=\epsilon_0+\epsilon_aZ^a-\mu^A\cG_A\; .
\end{equation}
For completeness, we also mention that the gauge kinetic function $f$ is given
by the dilaton, $f=s$, at order $\ap$, as expected for heterotic
compactifications.
 
Let us now come back to some of the questions raised in the introduction. We
have seen that the low-energy theory can be written in an $E_6$ invariant way
due to a suitable choice of the gauge connection and a related definition of
the matter fields $C$ and $D$. With the effective theory at hand, we should
now re-assess the question of what the low-energy gauge group actually is in
the light of a possible spontaneous symmetry breaking in the effective theory.
It is clear from the above results that for all choices of torsion and flux
parameters, there exists a supersymmetric vacuum at $C=D=0$ where the $E_6$
gauge group is, of course, unbroken. The existence of this vacuum means that
our choice of gauge bundle was sensible and has provided a suitable background
around which to consider fluctuations. When the $CD$ term in the
superpotential is present (that is, for generalised half-flat manifolds but
not for half-flat mirror manifolds) there is also the possibility of a
supersymmetric vacuum with $C,D\neq 0$ and of the order of the torsion
parameters. The gauge symmetry in this vacuum is presumably broken to ${\rm
  SO}(10)$ or even smaller. However, given that the torsion parameters are
presumably quantised in string units the $C$ and $D$ VEVs would be rather
large and it is doubtful if this vacuum can be considered as consistent in a
theory derived as an expansion in the matter fields. We will investigate this
question in detail in a forthcoming publication~\cite{dCLM}.  For the further
discussion, let us focus on the $E_6$ preserving vacuum at $C=D=0$.  
For generalised half-flat manifolds, when the term $CD$ in the superpotential
is present, some or all of the vector-like ${\bf 27}$, ${\bf\overline{27}}$
pairs receive a large mass and will be removed from the low-energy spectrum.  
This is an explicit realisation of the
usual lore by which vector-like pairs of matter fields are removed and, at low
energy, one remains with a net number of families given by the Euler number
$|\chi |/2=|h^{1,1}-h^{2,1}|$.  The above results open up various avenues for
exploring the phenomenology of the heterotic string on manifolds with ${\rm
  SU}(3)$ structure, including the question of heterotic moduli
stabilisation~\cite{dCGLM}, the precise nature of the family anti-family
pairing and, if supersymmetry breaking vacua are found, the computation of
soft masses and parameters. We hope to report on these issues in a future
publication~\cite{dCLM}.

\vspace{1cm}
\noindent
{\large\bf Acknowledgments} It is a pleasure to thank Kang-Sin Choi, Takayuki
Hirayama, Albrecht Klemm, Hans-Peter Nilles and Ashoke Sen for helpful
discussions and comments. A.~L.~is supported by the EC 6th Framework Programme
MRTN-CT-2004-503369. The work of A.~M.~is partially supported by the
European Union 6th framework program MRTN-CT-2004-503069 "Quest for
unification", MRTN-CT-2004-005104 "ForcesUniverse", MRTN-CT-2006-035863
"UniverseNet" and SFB-Transregio 33 "The Dark Universe" by Deutsche
Forschungsgemeinschaft (DFG).

\pagebreak


\vskip 1cm
\appendix{\noindent\Large \bf Appendix}
\renewcommand{\theequation}{\Alph{section}.\arabic{equation}}
\setcounter{equation}{0}


\section{Conventions and notations}
\label{conv}

In this appendix we present our conventions and formulae which we use
throughout the paper. 

\subsection{General conventions}

We denote real indices on the Calabi--Yau manifold by $m,n,\ldots = 1, \ldots,
6$, holomorphic ones by $\ax, \bx ,\ldots = 1, 2, 3$ and anti-holomorphic ones
by $\bar \ax, \bar \bx,\ldots = 1,2,3$.  Tangent space indices are referred to
with the same symbols as above, but with an additional tilde
underneath, so for example we use $\um$ for a real tangent space index.

Indices $i, j,\ldots =1 , \ldots , h^{1,1}$ and $a, b,\ldots = 1, \ldots,
h^{2,1}$ label objects on the moduli spaces of K\"ahler and complex structure
deformations, respectively.  We shall also use the capitalised versions of
these indices to label projective coordinates on these spaces, that is, for
example $A, B,\ldots = 0, 1, \ldots, h^{2,1}$ for the projective complex
structure moduli space.

Finally we use capital letters from the middle of the alphabet $M,N, ...$ for
the quantities which transform under $\rep{27}$ of $E_6$.

Where possible we shall use form notation. We use the
following conventions:
\begin{itemize}
\item We define a $p$-form as $F_p = \frac1{p!} F_{m_1 ... m_p} dx^{m_1}
  \wedge ... \wedge dx^{m_p}$.
\item the exterior product of a $p$- and a $q$-form is defined as \\[2mm]
  $F_p \wedge G_q = \frac{1}{p! \, q!} F_{m_1 ... m_p} G_{m_{p+1} ... m_{p+q}}
  dx^{m_1}\wedge ... \wedge dx^{m_{p+q}} $ which implies the component relation
  $(F_p \wedge G_q)_{m_1 ... m_{p+q}} = \frac{(p+q)!}{p! \, q!} 
  F_{[m_1 ... m_p} G_{m_{p+1} ... m_{p+q}]}$, where the antisymmetrisation is
  always understood to be of unit norm.
\item we define the Hogde star $*$ such that $F_p \wedge * F_p = \frac{1}{p!}
  F_{m_1 ... m_p} F^{m_1 ...m_p}$.
\end{itemize}

\subsection{Conventions for Calabi--Yau manifolds}
\label{CYconv}

We now collect some equations and conventions in relation to Calabi-Yau moduli
spaces.  They will be applied identically to the ${\rm SU}(3)$ structure
manifolds in sections \ref{sec:hf} and \ref{sec:genhf}, which are the main
subject of this paper.

We denote by $J$ the K\"ahler form and by $\Omega$ the holomorphic $(3,0)$
form on the Calabi--Yau manifold $X$. We choose a basis $(\ox_i )$ of harmonic
$(1,1)$ forms for the second cohomology group $H^{1,1}(X)$ and also introduce
the dual $(2,2)$ forms $\tox^i$. Further, we require a symplectic
basis $(\alpha_A, \beta^B)$ of the third cohomology.  These forms satisfy the
standard normalisation integrals
\begin{eqnarray}
  \label{norm2}
  \int \ox_i \wedge \tox^j & = & \delta_i^j \; , \\
  \label{norm3}
  \int \alpha_A \wedge \beta^B & = & \delta_A^B \; , \quad \int \alpha_A \wedge
  \alpha_B = \int \beta^A \wedge \beta^B = 0 \; .
\end{eqnarray}

The moduli space is parameterized by deformations of the K\"ahler form $J$ and
of the holomorphic $(3,0)$ form $\Omega$ which we expand as
\begin{eqnarray}
  \label{Kms}
  J & = & v^i \omega_i \; , \\
  \label{csms}
  \Omega & = & {\cal Z}^A \alpha_A - \cG_A \beta^A \; .
\end{eqnarray}
Here, $v^i$ denote the K\"ahler moduli and ${\cal Z}^A$ are projective
coordinates on the complex structure moduli space. Further, $\cG_A$ denote the
first derivatives of the prepotential $\cG$. The complex structure moduli are
given by
\begin{equation}
  \label{csmoduli}
  z^a = \frac{{\cal Z}^a}{{\cal Z}^0} \; ,
\end{equation}
and, for convenience,  we adopt the convention that  ${\cal Z}^0=1$.

The metrics on these moduli spaces can be written as 
\begin{eqnarray}
  \label{g11}
  g_{ij} & = & \frac{1}{4 \cK} \int \ox_i \wedge * \ox_j \; ,\\
  \label{g21}
  g_{a \bar b} & = & - \frac{1}{\int \Omega \wedge \bar \Omega} \int \chi_a
  \wedge \bar \chi_{\bar b} \; ,
\end{eqnarray}
where $\chi_a$ form a basis for the $(2,1)$ harmonic forms. Their relation to
the above symplectic basis $(\ax_A,\bx^B)$ is encoded in Kodaira's formula
\begin{equation}
  \label{kodaira}
  \frac{\partial\Omega}{\partial z^a}=-K_a\Omega+\chi_a\; ,
\end{equation}
where, $K_a$ denotes the derivative of the complex structure K\"ahler
potential, is given in \eqref{Kz}. The
inverse relations are somewhat more complicated to write down. They can be found
in the literature, for example in Appendix A of Ref.~ \cite{GLM} which follows the same conventions
as the present paper. Here we shall only give the formulae
for the $(1,2)$ parts
\begin{eqnarray}
  \label{bchi}
  (\bx^0)_{1,2} & = & - \frac{1}{\int \Omega \wedge \bar \Omega}  K_{b}
  g^{b \bar a} \bar \chi_{\bar a} \; , \nn \\
  (\bx^a)_{1,2} & = & - \frac{1}{\int \Omega \wedge \bar \Omega} \left(g^{a
  \bar b} + z^a K_c g^{c \bar b} \right)  \bar \chi_{\bar b}
  \; , \\
  (\ax_A)_{1,2} & = & - \frac{1}{\int \Omega \wedge \bar \Omega} g^{a \bar b}
  \big (\cG_{Aa} + K_a \cG_A \big) \bar \chi_{\bar b} \; , \nn
\end{eqnarray}
which are needed for various calculations throughout the paper.
As these forms are real, the $(2,1)$ parts can be simply obtained by
complex conjugation.

Since $\Omega$ is a $(3,0)$ form on a (almost) complex three-dimensional
manifold, it should be proportional to the complex $\epsilon$ symbol. We
write 
\begin{equation}
  \label{Oe}
  \Omega_{\ax \bx \cx} = ||\Omega|| \epsilon_{\ax \bx \cx} \; ,
\end{equation}
where the norm of $\Omega$ is defined as
\begin{equation}
  \label{normO}
  ||\Omega||^2 = \frac16 \Omega_{\ax \bx \cx} \bar \Omega^{\ax \bx \cx} =
  \frac{i}{\cK} \int \Omega \wedge \bar \Omega \; ,
\end{equation}
and $\cK$ denotes the volume of the Calabi--Yau manifold
\begin{equation}
  \label{vol}
  \cK = \frac16 \int J \wedge J \wedge J \; .
\end{equation}
Moreover, we use the conventions for the complex indices that $\Omega$ (and
thus $\epsilon$) is imaginary anti-self-dual (IASD)
\begin{equation}
  \label{starO}
  * \Omega = -i \Omega\; ,
\end{equation}
while the $(2,1)$ forms $\chi$ are imaginary self-dual (ISD)
\begin{equation}
  \label{starchi}
  * \chi = i \chi\; .
\end{equation}

We will frequently use the isomorphism between the space $H^{2,1}(X)$ of
$(2,1)$ harmonic forms and the space $H^{0,1}(X,TX)$ of $(0,1)$ harmonic forms
with values in the holomorphic tangent bundle whose elements we denote by
$\eta_a$. Explicitly, this isomorphism can be written as
\begin{equation}
  \label{eta}
  (\eta_a)_{\bar \ax}{}^\ax = \frac{1}{2 ||\Omega||^2} (\chi_a)_{\bx
  \cx}{}^{\ax} \bar \Omega_{\bar \ax}{}^{\bx \cx} \; ,
\end{equation}
In terms of the forms $\eta$, the metric on the moduli space of complex
structure deformations can be expressed as
\begin{equation}
  \label{g21eta}
  g_{a \bar b} = \frac{1}{\cK} \int (\eta_a)^{\ax \bx} (\bar \eta_{\bar
  b})_{\ax \bx} \; .
\end{equation}
Here, as in the rest of the paper, we have suppressed the measure $\sqrt{g}$
in the integral. 

The K\"ahler deformations $v^i$ can be viewed as imaginary parts of the
complexified fields $t^i$. Written in terms of these complexified fields, the
metric \eqref{g11} is K\"ahler with associated K\"ahler potential
\begin{equation}
  \label{Kt}
  K_K(t) = - \ln{\cK}  = - ln \left(\frac{1}{6} \cK_{ijk} v^i v^j v^k \right) 
  = - \ln \left[ \frac{i}{48} \cK_{ijk} (t^i - \bar t^i) (t^j - \bar t^j) (t^k
  - \bar t^k) \right ] \; .
\end{equation}
Here we have used Eq.~\eqref{vol} for the volume and the expansion
\eqref{Kms} of $J$. The triple intersection numbers $\cK_{ijk}$ are given by
\begin{equation}
  \label{tin}
  \cK_{ijk} = \int \ox_i \wedge \ox_j \wedge \ox_k \; .
\end{equation}

Similarly, the complex structure moduli space metric \eqref{g21} is also
K\"ahler with associated K\"ahler potential
\begin{equation}
  \label{Kz}
  K_{\rm cs}(z) = - \ln{i \int \Omega \wedge \bar \Omega} \; .
\end{equation}
This K\"ahler potential can be expressed explicitely in terms of the complex
structure moduli and the prepotential $\cG$ using Eq.~\eqref{csms}. In the
large complex structure limit the prepotential takes the form
\begin{equation}
  {\cal G}=-\frac{1}{6} \frac{{\tilde\cK}_{abc}{\cal Z}^a{\cal Z}^b{\cal
  Z}^c}{{\cal Z}^0}\; , 
  \label{csprep}
\end{equation}
with 
\begin{equation}
  \label{mtin}
  \tilde \cK_{abc} = -i \int (\eta_a)_{\bar \ax}{}^{\ax} (\eta_b)_{\bar
  \bx}{}^\bx (\eta_c)_{\bar \cx}{}^\cx \Omega_{\ax \bx \cx} \Omega^{\bar \ax
  \bar \bx \bar \cx} \; .
\end{equation}
being -- up to a constant normalisation -- the intersection numbers of the
Calabi-Yau manifold mirror to $X$. In this case, the complex structure 
K\"ahler potential $K_{\rm cs}$ is of the same form as the one for the K\"ahler
moduli~\eqref{Kt}, that is, 
\begin{equation}
  \label{Kcs1}
  K_{\rm cs}(t) = - \ln \left[ \frac{i}{48} \tilde\cK_{abc} (z^a - \bar z^a)
  (z^b - \bar z^b) (z^c 
  - \bar z^c) \right ] \; .
\end{equation}

\subsection{Commutators and traces}
\label{E8}

In this subsection we present our conventions for $E_8$ generators with
respect to the maximal subgroup $SU(3) \times E_6$. We split the $E_8$
generators into four groups, $S_{\ax \bar \bx}$, $T^x$, $T_{\ax P}$ and
$\bar{T}_{\bar\ax\bar P}$, in line with the decomposition
\begin{equation}
  \label{e8}
  (\rep{248}) = \underbrace{(\rep{8},\rep{1})}_{S_{\ax \bar \bx}} \oplus
  \underbrace{(\rep{1}, \rep{78})}_{T^x}  \oplus \underbrace{(\rep{3},
  \rep{27})}_{T_{\ax P}} \oplus \underbrace{(\rep{\bar 3},
  \rep{\overline{27}})}_{\bar T_{\bar \ax \bar P}} \; .
\end{equation}
of the adjoint of $E_8$ under $SU(3) \times E_6$.  Note that the index $P$ is
used to label objects which transform as a $\rep{27}$ under $E_6$ . The
matrices $S_{\ax \bar \bx}$ in the adjoint of ${\rm SU}(3)$ are subject to the
constraint $S_\ax{}^\ax =0$.

With these conventions the $E_8$ commutation relations can be written as
\begin{eqnarray}
  \label{cr327}
  \big[ T_{\ax P} , T_{\bx R} \big] & = &\epsilon_{\ax \bx}{}^{\bar \cx}
  j_{PR}{}^{\bar S} \bar T_{\bar \cx \bar S} \; , \\[.2cm]
  \label{cr327b}
  \big[ T_{\ax P} , \bar T_{\bar \bx \bar R} \big] & = & g_{\ax \bar \bx}
  k^x_{P \bar R} + g_{P \bar R} S_{\ax \bar \bx} \; ,\\[.1cm]
  \label{cr8327}
  \big[ S_{\bx \bar \cx} , T_{\ax P} \big] & = & - g_{\ax \cx} T_{\bx P} +
  \frac13 g_{\bx \bar \cx} T_{\ax P} \; , \\[.1cm]
  \label{cr2778}
  \big[T_{\ax P} , T^x \big] &=& - k^x_P{}^R T_{\ax R} \; .
\end{eqnarray}
Note that the $j_{PRS}$ is the fully symmetric, cubic invariant of $E_6$.
One can easily show that $-k^x_P{}^R$ are the components of the $E_6$
generators in the $\rep{27}$ representation. The $E_8$ Jacobi identity implies
that 
\begin{equation}
  \label{jid}
  j_{PR}{}^{\bar S} k^x_{S \bar S} + j_{RS}{}^{\bar S} k^x_{P \bar S} +
  j_{SP}{}^{\bar S} k^x_{R \bar S} = 0 \; .
\end{equation}

Finally we use the following normalisation for the $(\rep{3}, \rep{27})$
generators 
\begin{equation}
  \label{tr327}
  {\rm tr}(T_{\ax P} \bar T_{\bar \bx \bar P}) = g_{\ax \bar \bx} g_{P \bar R}
  \; . 
\end{equation}
Combining the last equation with Eq.~\eqref{cr327} one finds the useful
formula 
\begin{equation}
  \label{jdef}
  {\rm tr} \big( T_{\ax P} T_{\bx R} T_{\cx S} \big) = \epsilon_{\ax \bx \cx}
  j_{PRS} \; .
\end{equation}
for the cubic $E_6$ invariants $j_{PRS}$.


\end{document}